\documentclass[a4paper,11pt]{article}
\pdfoutput=1
\usepackage{amssymb,amsmath,amsfonts,makeidx,placeins,pbox,multirow}
\usepackage{graphicx,rotate,subcaption,color,slashed,caption,epstopdf,verbatim}
\usepackage{longtable,tabu}
\usepackage{array}
\usepackage{pstricks}
\usepackage{color}
\usepackage{amsmath}
\usepackage{mathtools}
\usepackage{amssymb}
\usepackage{bbold}
\usepackage[normalem]{ulem}
\usepackage{caption}
\usepackage{subcaption}
\usepackage{dsfont}

\newcolumntype{P}[1]{>{\centering\arraybackslash}p{#1}}
\usepackage[colorlinks=true,
            linkcolor=magenta,
            urlcolor=blue,
            citecolor=blue]{hyperref}

\evensidemargin=\oddsidemargin
\newcommand{\beq}{\begin{equation}}
\newcommand{\eeq}{\end{equation}}
\newcommand{\bea}{\begin{eqnarray}}
\newcommand{\eea}{\end{eqnarray}}

\def\to {\rightarrow}

\def\bea {\begin{eqnarray}}
\def\eea {\end{eqnarray}}

\def\barr{\begin{array}}
\def\earr{\end{array}}
\def\to{\end{rightarrow}}

\def\gev{\ensuremath{\mathrm{Ge\kern -0.1em V}}}

\usepackage{color}
\definecolor{amber}{rgb}{1,0.45,0}

\usepackage{tikz}
\usetikzlibrary{positioning,arrows}
\usetikzlibrary{decorations.pathmorphing}
\usetikzlibrary{decorations.markings}

\begin{document}
\tikzset{
vector/.style={decorate, decoration={snake,amplitude=.6mm}, draw=red},
scalar/.style={dashed, draw=blue},
fermion/.style={draw=black, postaction={decorate},
        decoration={markings,mark=at position .55 with {\arrow[draw=black]{>}}}}
}
\begin{center}
{\Large \bf Attenuation of boosted dark matter in two component dark matter scenario}\\
\vspace{0.3cm}
\vspace*{0.2cm} {\sf Nilanjana Kumar\footnote{nilanjana.kumar@gmail.com}, Gaadha Lekshmi \footnote{gaadha15@gmail.com}} \\
\vspace{6pt} {\small } {\em Centre for Cosmology and Science Popularization, SGT University, \\Gurugram (Haryana)-122505, India}
\normalsize
\end{center}

\begin{abstract}

Boosted dark matter constitutes a small fraction of the total dark matter in the Universe, with mass ranging from eV to MeV 
and often exhibiting (semi)relativistic velocity. Hence the likelihood of detecting boosted dark matter in Earth-based direct detection 
experiments is relatively high. There is more than one explanation for the origin of the boosted dark matter including 
the two-component dark matter models where the heavier dark matter species(dominant) annihilates to nearly monoenergetic light dark matter 
particles (subdominant) in the galactic halo. If the dominant dark matter species is heavier (MeV-GeV),
the subdominant light dark matter achieves (semi)relativistic velocity or {\it boost}.
These boosted dark matter particles suffer from scattering with electrons 
and nuclei while crossing the atmosphere and the Earth's crust before reaching underground experiments and hence the kinetic energy of the dark matter is attenuated. 
In the two-component dark matter framework, we examine how the boost of the dark matter influences the attenuation of kinetic energy across a broad spectrum of dark matter masses. 
We perform a detailed study at various DM-electron and DM-nucleus cross sections including the effect of nuclear form factor and elastic and inelastic scattering (for large kinetic energy).  
For a 10 MeV boosted dark matter with boost $\sim$ 10-100, the effect of DM-electron scattering is found to be severe than the DM-nucleus 
scattering (with form factor) if DM-nucleon scattering cross section is $10^{-29}$cm$^2$. 
We also show how the peak position of the boosted dark matter flux shifts due to the attenuation of its kinetic energy.

\end{abstract}
\section{INTRODUCTION}
The observational evidence of dark matter (DM) \cite{Bertone:2004pz,Planck:2018vyg} in the Universe and its gravitational effects are well explained by the idea of cold dark matter (CDM). It is a popular assumption that dark matter interacts very {\it weakly} with ordinary matter which is composed of Standard Model (SM) particles and its thermal relic abundance \cite{Planck:2018vyg} is set by its direct couplings to the standard model particles. Thus it is very challenging to detect the dark matter particles. Direct detection experiments which are in deep underground, observe the scattering of the DM particle with the target element, and measure the recoil energy of electron or the nucleus. However, the average velocity of the CDM ($\sim$ 220 km/s) in the galactic halo restricts the amount of recoil energy deposited in the detectors. Experiments such as XENONnT \cite{XENON:2023cxc}, LUX-ZEPLIN(LZ) \cite{LZ:2022lsv}, DEAP-3600 \cite{Lai:2023qub}, CRESST \cite{CRESST:2019jnq}, SuperCDMS \cite{SuperCDMS:2020ymb}
and others \cite{EDELWEISS:2020fxc,SENSEI:2020dpa} have put strong limits on the mass of CDM. 
The stringent limits put by such experiments on the DM-nucleon spin-independent (DM-electron) cross section is $\mathcal{O}(10^{-48}$cm$^2$) ($\mathcal{O}(10^{-41}$cm$^2$)) for dark matter mass $\mathcal{O}$($10$ GeV) \cite{LZ:2022lsv} ($\mathcal{O}$($100$ MeV) \cite{XENON:2023cxc}). However, most of the current direct detection experiments are not sensitive to the dark matter masses below 100 MeV. Light dark matter particles are predicted in many Beyond Standard Model (BSM) theories \cite{Fayet:2004bw,Boehm:2003hm,Boehm:2002yz}. 

If the light dark matter particles achieve relativistic velocity at the present Universe, the recoil energy of electrons or the nucleons 
will be large at the detectors, forcing the light dark matter particles in the detectable range of the direct detection experiments. 
These light dark matter particles, traveling with relativistic speed are termed as boosted dark matter (BDM). 
There are more than one natural explanation of the boosted dark matter production: (1) Interaction with cosmic rays (CR) or high energy neutrinos \cite{Xia:2021vbz,Jho:2021rmn,Yin:2018yjn}, 
(2) Evaporation from primordial black holes (PBH) \cite{Chao:2021orr}, (3) DSNB and Blazar boosted dark matter \cite{DeRomeri:2023ytt,Das:2024ghw,Granelli:2022ysi},
(4) Annihilation of the heavy (dominant) DM in the galactic halo at present time \cite{Basu:2023wgo,Li:2023fzv} and others \cite{Herrera:2021puj,Herrera:2023fpq}.
In all these cases, the boosted dark matter is the subdominant component of the total DM relic and hence the flux of the boosted dark matter is comparatively small.

Due to the small flux of the BDM, large volume detectors are preferred for BDM detection.
All the Earth based large volume detectors are situated deep underground. Experiments such as XENONnT \cite{XENON:2023cxc}, LUX-ZEPLIN(LZ) \cite{LZ:2022lsv}, DarkSide-20k \cite{DarkSide-20k:2017zyg} and PandaX \cite{PandaX:2014mem} are at distance of 1.4 km-2.4 km from the surface of Earth with target material volume in the range of 2-20 metric tons. The BDM particles travel through the atmosphere and Earth's crust before reaching the detectors. The BDM particles collide with the particles in the atmosphere and the elements inside the Earth's crust and its initial kinetic energy is attenuated. As a result, the BDM flux also suffers from attenuation.
The attenuation effect due to the atmosphere is very less compared to the attenuation due to the Earth's crust \cite{Xia:2021vbz}. 
Effect of attenuation on the boosted dark matter has been studied recently in the context of CR or blazer boosted dark matter and boosted dark matter from evaporation of PBH with great importance \cite{Chen:2021ifo,Herbermann:2024kcy}. It has been shown that if the effect of attenuation is considered, the experimental predictions of the upper and lower exclusion limits change at LZ and XENON experiments \cite{DeRomeri:2023ytt,Xia:2021vbz}. For example, if the DM gets boosted by collision with DSNB, DM mass is between 0.1 MeV -1000 MeV, and the kinetic energy is less than 10 MeV, then scattering cross sections larger than $\geq \mathcal{O}(10^{-28})$ cm$^2$
are disfavored \cite{DeRomeri:2023ytt}. Whereas, Ref:\cite{Xia:2021vbz} shows that for CR boosted DM, cross section larger than $\mathcal{O}(10^{-28})$ cm$^2$ are excluded if DM mass is $\geq \mathcal{O}(1)$ MeV and if the kinetic energy $\geq \mathcal{O}(1)$ GeV. These exclusion limits on the cross section not only depends on the mass of the dark matter, and kinetic energy, the distribution of kinetic energy, but also on the flux of BDM and its source.

In this paper, we choose a working model of two component dark matter, where the annihilation of the dominant DM species $A$ produces the boosted dark matter $B$. Boosted dark matter in two component scenarios are studied in Ref:\cite{Basu:2023wgo,Li:2023fzv,Agashe:2014yua,Borah:2021yek,Borah:2021jzu} and the interactions are modeled in such a way that there exist a very small annihilation rate for $AA\rightarrow BB$ at present day which is responsible for producing relatively small amount of highly energetic dark matter flux for $B$. 
The annihilation $AA\rightarrow BB$ produces two nearly mono-energetic boosted dark matter $B$ with energy ($T_B$) close to the mass of $A$ ($T_B\sim m_A$), hence the boosted dark matter flux can be approximated as a normal distribution with small width ($\sigma$), and in the limit $\sigma \rightarrow 0$, it tends to a Dirac delta distribution. Thus the flux of the BDM in this case is nearly mono-energetic as compared to the boosted dark matter flux in CR or DSNB boosted dark matter scenario.  
Also, the kinetic energy of the BDM and the boost are not independent quantities but are functions of the dark matter masses.
Another novelty of these models is that the position of the peak of BDM flux is at $T_B= m_A$. Hence detection of the boosted dark matter $B$ gives us insight about the nonboosted dark matter component $A$ as well. 

In some scenarios the boost of the DM particle ($B$) has turned out to be very large ($\geq10^4$)\cite{Basu:2023wgo}. The boost of the DM candidate enhances its detectability in experiments, and for such high boost, the attenuation of kinetic energy ($T_B$) due to the atmosphere and Earth's crust is negligible.
In this paper we study the attenuation of the boosted dark matter's kinetic energy due to the collision with both electron and the nucleons for two scenarios:
(1)Boost($\geq{O}(10^2)$): When the boost and the kinetic energy of the BDM is very high, and 
(2)Boost($\leq{O}(10^2)$): When the boost and the kinetic energy of the BDM is moderate. 
We follow the analytical approach which is based on one dimensional collision approximation \cite{Starkman:1990nj,Kouvaris:2014lpa}. 
Previously, it has been shown that the inclusion of nuclear form factor affects the attenuation of the kinetic energy \cite{Xia:2021vbz}.
In this paper, the effect of including the nuclear form factor \cite{Helm:1956zz} in the DM-nucleon scattering is studied in detail for the above scenarios.
If the kinetic energy of the boosted dark matter is smaller than 1 GeV, the scattering is largely elastic. However, Ref. \cite{Su:2022wpj,Su:2023zgr,PandaX:2023tfq,Alvey:2022pad} have shown that, for kinetic energy of the BDM greater than 1 GeV the effect of inelastic scattering (IES) dominates over elastic scattering (ES). At such high energy, due to large energy transfer, the BDM particle interacts directly with the constituent nucleons. Hence, at large kinetic energies, both elastic and inelastic scattering will contribute.

The DM-electron scattering is also not negligible for the MeV scale boosted dark matter. In this paper we consider both the scattering process in detail and their corelation with the dark matter masses and kinetic energy. We also show how the attenuation depends on the value of DM-nucleon($\sigma_{Bn}$) and DM-electron($\sigma_{Be}$) scattering cross sections. 
The cross section is assumed to be constant and independent of the underlying model so that the results can be translated to any other two component boosted dark matter scenario.
We also present the attenuation of the BDM flux and the shift in the peak position of the BDM flux as a function of DM masses $m_A$, $m_B$ and the scattering cross section/s.
For large attenuation, the position of the peak in the attenuated BDM flux does not always predict the correct mass of the dominant dark matter ($A$).

In Section 2, we discuss the galactic flux of the boosted dark matter for a two component dark matter model and its variation with the boost and different dark matter density profiles and annihilation cross section. Section 3 focuses on the attenuation of boosted dark matter as it traverses through the atmosphere and the Earth's crust. First (Section 3.1) we consider the nuclear scattering with form factor $=1$ and then we extend this analysis to cases where the nuclear form factor is not equal to one (Section 3.2). The effect of the nuclear form factor and its variation with the kinetic energy of the boosted dark matter particle is studied in detail. In Section 3.3, we study the scattering of boosted dark matter with electrons, exploring its effect on the kinetic energy and flux of the BDM. In Section 3.4 we present the combined result. Finally, in Section 4 we conclude with our findings.

\section{GALACTIC FLUX OF BOOSTED DARK MATTER }
In a two component dark matter model, the interaction among the DM candidates and/or with Standard Model (SM) particles depend on the 
details of model building. 
The interaction between two DM particles can be modeled by portal type interactions, mediated by Standard Model particles such as Higgs boson or BSM particles such as a dark photon \cite{Agashe:2014yua}, heavy or light Higgs \cite{Bhattacharya:2013hva}, or gauge bosons \cite{Guha:2024mjr}. 
However, we adopt a model independent analysis of two component dark matter scenario in this paper. 

Let us consider a working model of two component dark matter. In this model the boosted dark matter ($B$) is produced today at the galactic halo via the annihilation of the dominant DM ($A$): $AA\rightarrow BB$.
If the dark matter $A$ is at rest then the centre of mass energy is $s\sim 4m_{A}^2$ and the annihilation cross section in the non relativistic limit can be approximated as, 
$\langle\sigma_{AA\rightarrow BB} ~v\rangle \sim \sigma_{AA\rightarrow BB}~v$. This annihilation cross section is a function of the model parameters in any 
two component dark matter model. The cross section is very large if there is a resonance effect due to the mass of the mediator particle \cite{Basu:2023wgo}. Whereas for models with mass less mediators (dark photon \cite{Agashe:2014yua}) such effects are absent. In this analysis we choose some benchmark values of the annihilation cross section to address the vast range of possible annihilation cross sections predicted in different literature.

\begin{figure}[htb!]
\begin{center}
\includegraphics[width=7cm,height=5cm]{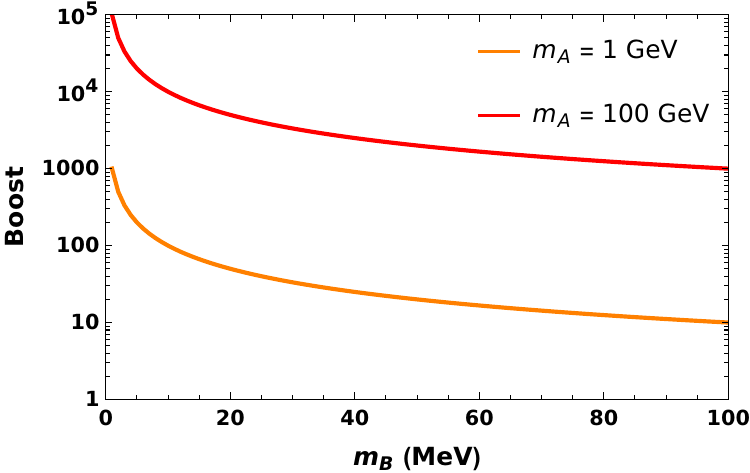}
\includegraphics[width=7cm,height=5cm]{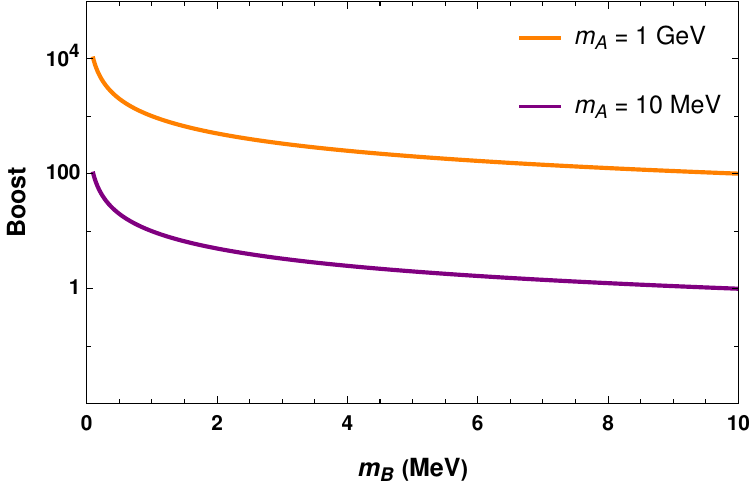}
\end{center}
\caption{Boost as a function of mass $m_B$ when $m_A\geq 1$ GeV (left) and $m_A\leq 1$ GeV (right).}
\label{Boost_plot}
\end{figure}

If there is significant mass difference among two DM candidates $A$ and $B$, $B$ achieves very high velocity. This is described in terms of {\it boost}($\gamma$) which is a function of the velocity ($v_B$) of the dark matter. Boost can be expressed as the ratio of the DM candidate masses.
\begin{align}
\gamma= \frac{1}{\sqrt{1-v_B^2}}\sim\frac{m_A}{m_B}
\end{align}
The kinetic energy of the boosted dark matter ($B$) is  
\begin{align}
T_B= \gamma m_B.
\end{align}
The masses of the dark matter particles in a particular model are constrained from the DM relic density measurement \cite{Planck:2018vyg} and other experimental observations. 
Here we consider two cases:  $m_A >$1 GeV and $m_A< 1 $ GeV and plot the boost as a function of $m_B$ in Fig.\ref{Boost_plot}. We found that depending on the mass of the dark matter $B$ the boost can be high ($\geq \mathcal{O}(10^2$)) or moderate ($\leq \mathcal{O}(10^2$)). When viewed from the perspective of the boosted dark matter particle detection, the detection sensitivity is found to be larger even if the dark matter particles are boosted by a factor of $\mathcal{O}(10^2)$ \cite{Li:2023fzv}.

The differential flux of the boosted dark matter $B$ from the galactic center is \cite{Agashe:2014yua},
\begin{align}
    \frac{d \phi}{d \Omega \, d T_B} = \frac{1}{4} \frac{r_{\text{Sun}}}{4 \pi} \left( \frac{\rho_{\text{local}}}{m_{A}}\right) ^2 J \,  \langle\sigma_{AA\rightarrow BB}~v\rangle \frac{d N_{B}}{dT_{B}},
\end{align}
where $r_\text{Sun}$ = 8.33 kpc is the distance from the sun to the GC and $\rho_{local}$ = 0.3 GeV/cm$^3$ is the local DM density. 
Since the $AA\rightarrow BB$ annihilation process produces two nearly monoenergetic boosted dark matter $B$ with energy close to the mass of $A$, the differential energy spectrum is expressed as,
\begin{align}\label{eq:energy_spectrum}
    \frac{d N_{B}}{d T_{B}} = 2 \, \delta(T_{B} - m_{A})
\end{align}
Note that, in this scenario the boosted dark matter particles have the same energy equal to $m_A$. Also, the 
mass of the boosted dark matter ($B$) has no effect in the galactic flux unlike the CR or DSNB boosted dark matter scenario, where the flux 
is highly dependent on the mass of the dark matter particle.

We also define the following dimensionless integral over the line of sight, which depends on the DM density profile. 
\begin{align}\label{eq: J term}
    J = \int_{\text{l.o.s}} \frac{d s }{r_{\text{Sun}}} \left( \frac{\rho(r(s,\theta))}{\rho_{\text{local}}} \right)^2 
\end{align}
The most commonly used density profile is the Navarro-Frenk-White (NFW) \cite{Navarro:1995iw} given by 
\begin{align}
    \rho_{NFW}(r) = \frac{\rho_{local}}{\frac{r}{R}(1+\frac{r}{R})^{2}}
\end{align}
where $r$ is the distance from the galactic center and R is the scale radius of Milky Way. This profile features a steep (cusp) at the central region of the galaxy that falls as $\rho(r) \propto \frac{1}{r} $. In addition to NFW, another ``cuspy'' profile is the Einasto profile \cite{1965TrAlm...5...87E}. On the other hand, the Burkert \cite{Burkert:1995yz} profile has a corelike structure, featuring a constant density at smaller radii that eventually falls off at larger radii. 
\begin{align}
    \rho_{Burkert}(r) = \frac{\rho_{local}}{(1+\frac{r}{R})(1+\frac{r^{2}}{R^{2}})}
\end{align}
For the different profiles, we use the interpolation function $J(\theta)$ provided in \cite{Cirelli:2010xx} and integrate it over the whole sky where $\theta$ ranges from 0 to $\pi$ and $\phi$ ranges from 0 to $2\pi$.
The values of $J$ for different profiles over the whole sky is derived and given in Table\ref{prof_table}.
\begin{table}[h!]
\centering
\begin{tabular}{|c|c|}
\hline
Profiles & $J$ (whole sky) \\
\hline
Einasto & 42.27 \\
\hline
 NFW & 34.58 \\
\hline
 Burkert & 19.17 \\
\hline
\end{tabular}
\caption{Values of $J$ for different DM profiles integrated over the whole sky.}
\label{prof_table}
\end{table}

\begin{figure}[htb!]
\begin{center}
\includegraphics[width=7cm,height=5cm]{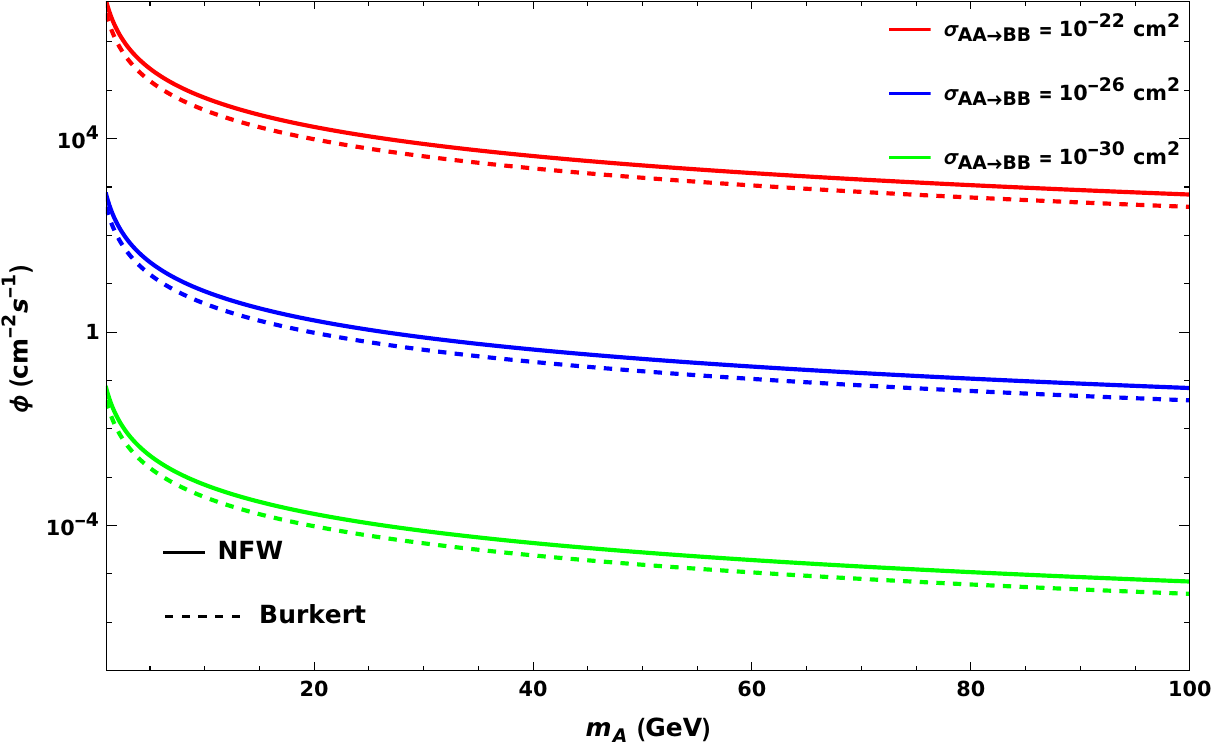}
\includegraphics[width=7cm,height=5cm]{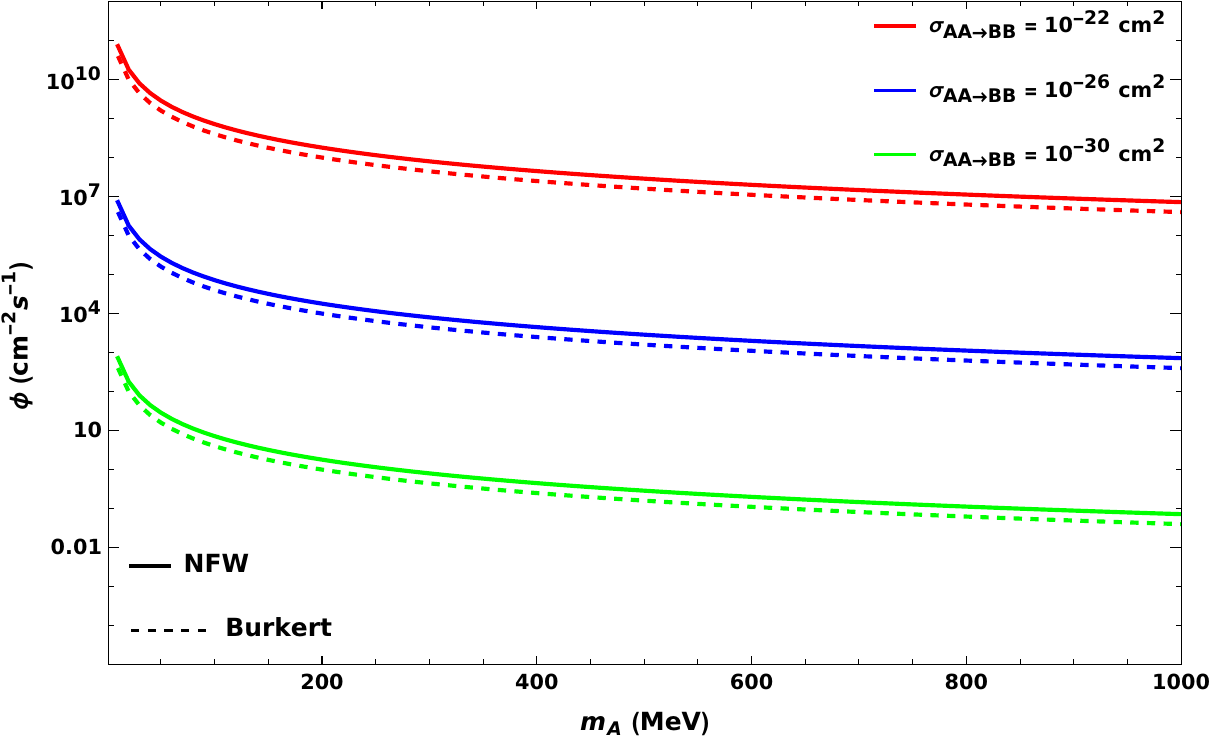}
\end{center}
\caption{Total BDM flux as a function of $m_A$ is shown for the monochromatic distribution assumed in Eq. (\ref{eq:energy_spectrum}) at different benchmark values of the annihilation cross section. The dominant dark matter masses are $m_A>1$ GeV (left) and $m_A<1$ GeV (right).}
\label{Total_flux}
\end{figure}

In Fig.\ref{Total_flux} we plot the total flux of the BDM by integrating Eq. (\ref{eq: J term}) over the whole sky. The total flux is directly proportional to the annihilation cross section and inversely proportional to $m_A^2$.
Hence, the flux increases for the larger values of the annihilation cross section. Also in the scenario $m_A <1$ GeV large amount of flux is produced. 
In Fig.\ref{Total_flux} we also show the comparison between the NFW and Burkert profiles and we found that choosing NFW profile always results in larger amount of flux.

\begin{figure}[htb!]
\begin{center}
\includegraphics[width=7cm,height=5cm]{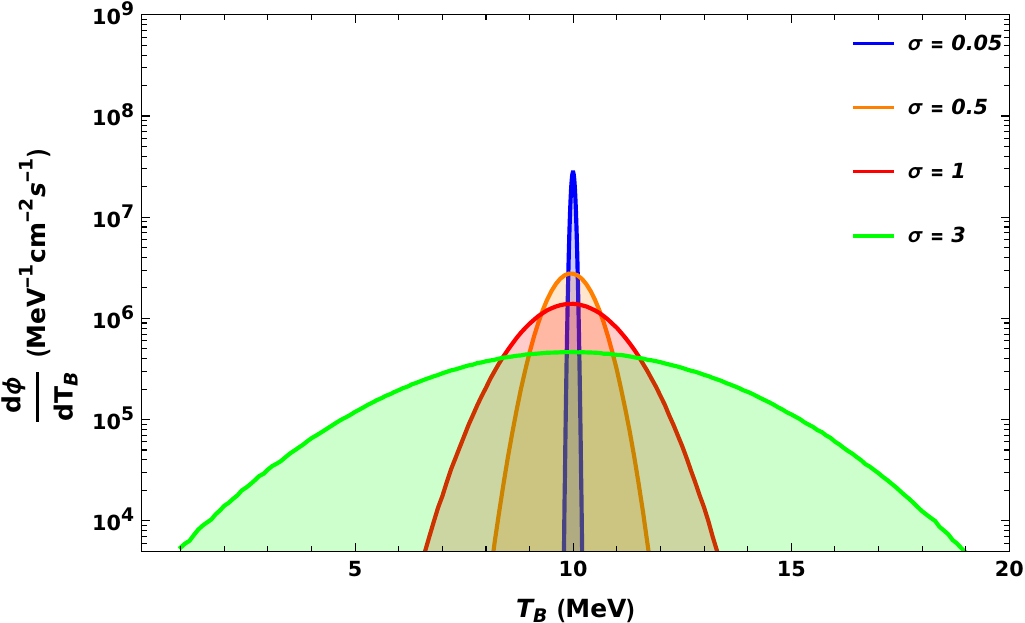}
\includegraphics[width=7cm,height=5cm]{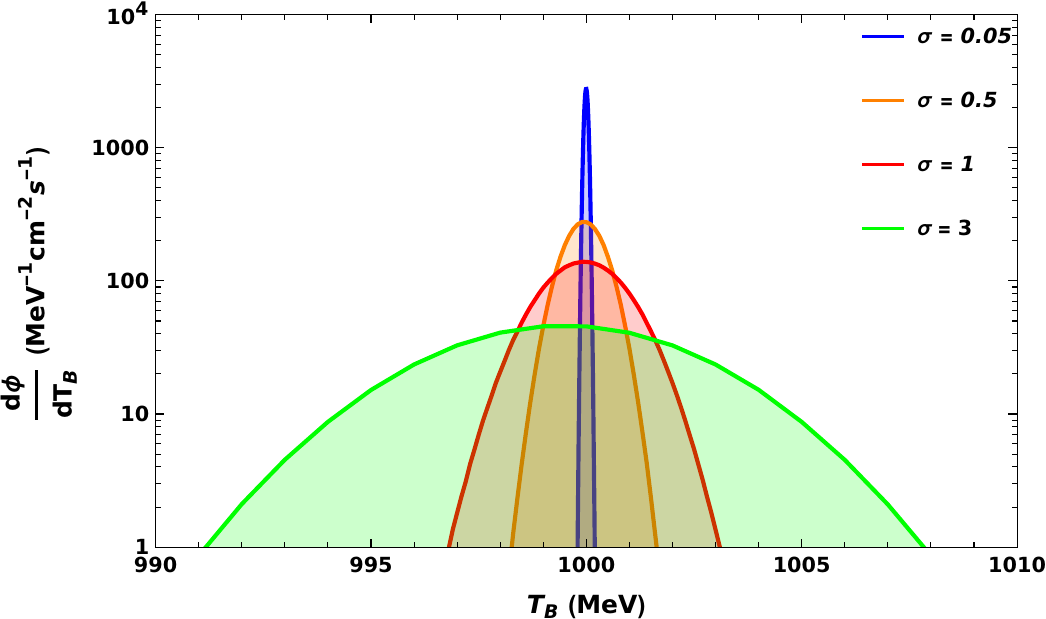}
\end{center}
\caption{Differential flux of the boosted dark matter as a function of its kinetic energy ($T_B$) at different width ($\sigma$) of the normal distribution. The production cross section is fixed at $\sigma_{AA\rightarrow BB}=10^{-26}$ cm$^2$.}
\label{diff_flux_clrd}
\end{figure}

Our discussion so far is based on the scenario that all boosted dark matter particles ($B$) have same energy ($=m_A$) due to the ideal condition assumed in Eq. (\ref{eq:energy_spectrum}), where the dark matter particle $A$ only couples with other dark matter candidate $B$ \cite{Agashe:2014yua}. Now, in some models, $A$ couples with the SM as well and hence the annihilation $AA \rightarrow SM SM $ is allowed. In this case, the flux of dark matter $B$ originating form the process $AA \rightarrow BB$ (still active but by a small amount in present Universe) may have a distribution over a range of energy \cite{Li:2023fzv}. This energy distribution of the boosted dark matter depend on the kinematics of the exact model and also the velocity distribution of the dark matter $A$ in the galaxy\cite{Bhattacharjee:2012xm}. 
Hence it is a reasonable assumption that the flux of the boosted dark matter follows a normal distribution over a range of kinetic energy, with $\sigma$ being the width of the distribution. In the limit $\sigma$ tends to zero, we get back the limiting case in Eq. (\ref{eq:energy_spectrum}). The energy distribution of the boosted dark matter can be wide or narrow depending on the values of $m_A$ that are allowed by the model of concern. Here we choose the width $\sigma\leq 3$ \footnote{For the DSNB and CR boosted dark matter, the energy distribution of flux is wide, its kinetic energy ranges from KeV to GeV \cite{DeRomeri:2023ytt,Xia:2021vbz}.}. In Fig.\ref{diff_flux_clrd} we show the differential flux $\frac{d \phi}{dT_{B}}$ for two different energy regions. The kinetic energy of the boosted DM has a peak at $T_B=m_A=10$ MeV in the left plot and at $T_B=m_A=1$ GeV in the right plot.
The differential flux is calculated by assuming different $\sigma$ as well.

\section{ATTENUATION OF THE BOOSTED DARK MATTER}
We have observed in the previous section that the kinetic energy of the  boosted dark matter is either very large ($\sim$ GeV) or moderate ($\sim$ MeV), depending on the mass of the annihilating dark matter mass ($m_A$). Hence there is a greater chance to observe the boosted DM in the direct detection experiments even if the mass of the boosted dark matter ($B$) is in the MeV range.
The flux of dark matter particles passes through the atmosphere and the crust of Earth before reaching the underground detection experiments. Hence the dark matter particle suffer from elastic collision with the electrons and nucleus of the atoms in their journey toward the detectors. As a result, the kinetic energy of the boosted dark matter reaching the detector is attenuated compared to its initial value. However, inelastic scattering have significant effect at large kinetic energies \cite{Su:2022wpj}.
Hence it is very crucial to study the attenuation of the kinetic energy and the flux of the dark matter before it reaches the underground experiments.

When the dark matter particle crosses the crust of the Earth, the energy loss is significant compared to the energy loss while crossing Earth's atmosphere \cite{Xia:2021vbz}. If the dark matter is highly boosted, then there is a lesser chance to suffer from collisions, and hence the attenuation of the kinetic energy is sometime neglected. In the following we show that the attenuation not only depends on the kinetic energy of the boosted dark matter, but also depends on the electron and nucleon scattering cross section and the mass of the boosted dark matter particles as well.

The rate of change of the kinetic energy of the boosted dark matter ($B$) with respect to the distance $z$ is 
\begin{equation}\label{eq:attenuation}
	\frac{d T_B}{d z} = - \sum_i
	n_i \int_0^{T_i^{\max}} \frac{d \sigma_{B i}}{d T_i} T_i d
	T_i,
\end{equation}
where, $i = e, N$ for electron or nucleus species respectively and $n_i$ is the number densities of nucleus species $N$ or electron in the Earth crust. $n_N$ for oxygen is calculated \footnote{Note that $n_N = \rho.f_i/m_i$ where $\rho$ is constant mass density of Earth's crust, $f_i$ is the mass fraction of the nuclei with mass $m_i$. For oxygen, $n_N = 1.514\times 10^{27}$ (MeV/cm$^3)\times 0.466 / 14900$ (MeV) \cite{Emken:2018run}.} to be $4.7 \times 10^{22}$ cm$^{-3}$ and for electron we assume a constant number density, $n_e = 8 \times 10^{23}$ cm$^{-3}$ \cite{Ema:2018bih}. $z$ is the distance traveled by the BDM particles from the point of impact on the Earth's surface to the location of the detector. $z$ can be expressed as\cite{DeRomeri:2023ytt}
\begin{equation}
    z = -(R_E- h_d)\cos\theta_z + \sqrt{R_E^2-(R_E - h_d)^2\sin^2\theta_z}
\end{equation}
where, $R_E$ is the radius of the Earth, $\theta_z$ is the detector’s zenith angle and $h_d$ is the depth at which the detector is located and where the zenith angle is zero. Throughout this work, we assume the case where $\theta_z$ = 0 (z = $h_d$).

The differential cross section for DM-electron/nucleon collision is expressed as
\begin{equation}
\frac{d \sigma_{B i}}{d T_i}=\frac {\sigma_{B i}}{T_i^{\max}(T_{B})}
\end{equation}
and the maximal recoil energy of the electron or the nucleon is found to be 
\begin{equation}\label{eq:kin_tb_max}
    T_i^{\max}(T_{B})= {\left[1
    + \frac{{(m_i-m_{B})}^2}{2m_i(T_{B}+2m_{B})}\right]}^{-1}T_{B} .
\end{equation}
\subsection{Nuclear scattering: Form factor $=1$}
Let us first discuss the scattering of the boosted dark matter with the nucleus first. The DM-nucleon cross section $\sigma_{Bn}$ depends on the mediator particle/s which communicates between the dark sector and the visible sector. 
The DM-nucleon cross section has two parts, one is spin dependent and the other is spin independent. The spin of the nucleus arises from the unpaired nucleon. Hence the spin dependent cross section is more important for the light nuclei \cite{Wu:2022jln}. However, the chemical component of Earth tells us that it is made up with moderately heavy elements $\sim 47$\% oxygen ($^{16}$O), $\sim 28$\% silicon ($^{16}$Si) and rest is mostly iron, calcium, potassium, sodium and magnesium \cite{Emken:2018run}. Thus we can neglect the spin dependent cross section while studying the scattering of the DM due to the abundance of the above elements on Earth. 

The spin independent part of the DM-nucleus cross section, $\sigma_{BN}$, is related to the atomic number ($A$) and DM-nucleon cross section, $\sigma_{Bn}$, as
\begin{equation}
\sigma_{B N}(q^2)=\frac{\mu_N^2}{\mu_n^2} A^2 \sigma_{Bn} F^2(q^2)
\end{equation}
where, $\mu_N$ is the DM-nucleus reduced mass and $\mu_n$ is the DM-nucleon reduced mass.
$A$ is the mass number of the nuclei and $q$ denotes the momentum transfer
$\sqrt{2m_N T_N}$, where $T_N$ is the nuclear recoil energy. Here, we assume the DM-nucleon scattering to be isospin conserving. The form factor $F(q^2)$ accounts for the finite size of the nucleus.
One may approximate it with Helm form factor ~\cite{Helm:1956zz,Lewin:1995rx}. For the heavier nuclei, Helm
form factor takes the following form:
\begin{equation}\label{eq:helmformfactor}
	F(q^2) = \frac{3j_1(qR_1)}{qR_1} e^{- \frac{1}{2} q^2 s^2} 
\end{equation}
where $j_1(x)$ is the spherical Bessel function of the first kind,
$R_1 = \sqrt{R_A^2-5s^2}$ with $R_A \approx 1.2 A^{1/3}~\text{fm}$,
and $s\approx 1~\text{fm}$.

Since the dominant chemical component of Earth's crust is oxygen,  $m_N\approx 15~\text{GeV}$, we consider the
case where $m_B, T_{B} << m_N$. 
In this limit we can write, 
\begin{align}\label{eq:dTdz}
	\frac{d T_B}{d
	z}=-\frac{1}{\ell(T_B)}\left(T_B+ \frac{T_B^2}{2
	m_B}\right).
\end{align}
Here $\ell$ is the mean free path for energy loss, given by, 
\begin{align}\label{1_by_ell}
\frac{1}{\ell(T_B)}=2 m_{B}\sum_N g_N(T_B) n_N \sigma_{B N}/m_N
\end{align}
The factor $g_N$ is calculated by integrating the following quantity that involves the nuclear form factor as 
\begin{equation}\label{gN}
    g_N(T_B)=\int^{T^{\text{max}}_N}_{0}
    F^2_N(q^2)\frac{2T_N}{{(T^{\text{max}}_N)}^2} dT_N.
\end{equation}
We first solve for the attenuation of kinetic energy in the limiting case, when $F_N(q^2) \rightarrow 1$. In this limit $g_N$ approaches unity. Hence,
\begin{align}\label{1byell}
\frac{1}{\ell}=2 m_{B}\sum_N n_N \sigma_{B N}/m_N
\end{align}
and the analytical solution for the kinetic energy of the boosted dark matter at a distance $z$ is,
\begin{equation}\label{eq:Tz_noformfac}
	T_B(z)=\frac{T_B(0)
	e^{-z/\ell}}{1+\frac{T_B(0)}{2
	m_B}\left(1-e^{-z/\ell}\right)}.
\end{equation}
\begin{figure}[htb!]
\begin{center}
\includegraphics[width=7cm,height=5cm]{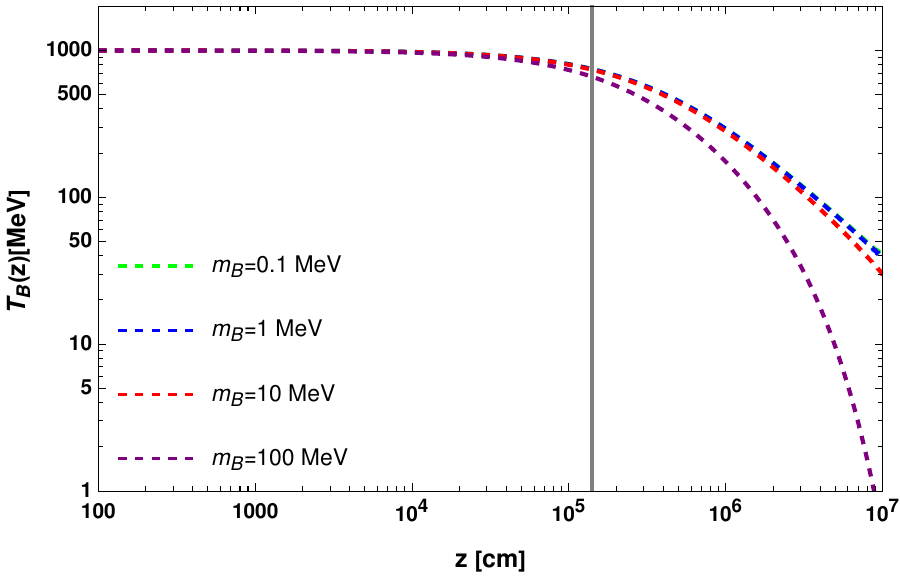}
\includegraphics[width=7cm,height=5cm]{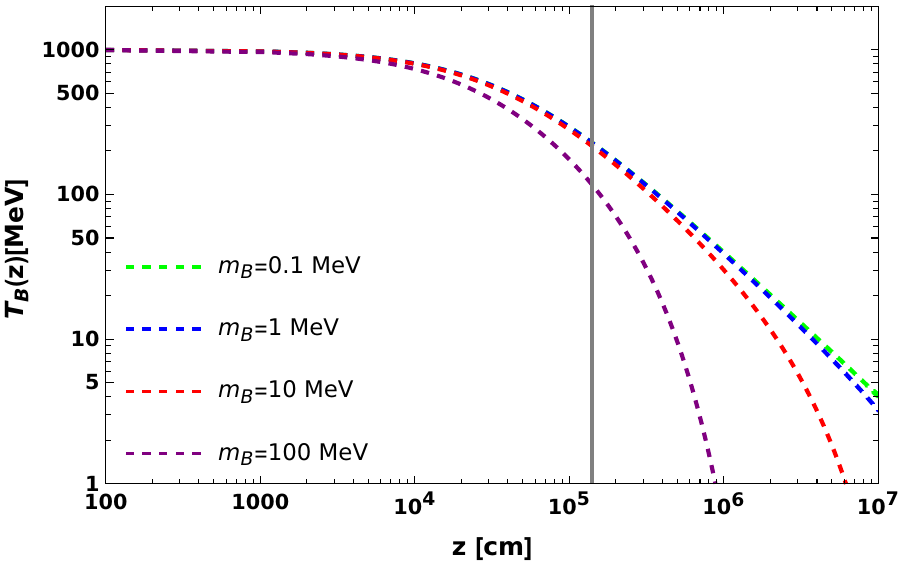}
\end{center}
\caption{Attenuation of the kinetic energy of boosted dark matter as a function of distance ($z$) is shown when $F_N^2(q^2)= 1$ and for different masses of the dark matter ($m_{B}$). $\sigma_{Bn} = 3\times10^{-30}$cm$^2$ (left) and $\sigma_{Bn} = 3\times10^{-29}$cm$^2$ (right). The gray vertical line shows the depth of XENON experiment, $z = 1.4$ km.}
\label{Attn_plot1}
\end{figure}
In Fig.\ref{Attn_plot1} we show the effect of attenuation of the kinetic energy for different DM masses with no form factor. Note that, even if the initial kinetic energy is same for all $m_B$, the boost is different( $\gamma=m_A/m_B\sim T_B/m_B$) in each case. Fig.\ref{Attn_plot1} shows that even if the DM-nucleon cross section is larger by one order of magnitude, the kinetic energy is more attenuated. Also, if the DM is very light, the kinetic energy suffers from less attenuation compared to the heavier DM when they have same initial kinetic energy.
\begin{figure}[htb!]
\begin{center}
\includegraphics[width=7cm,height=5cm]{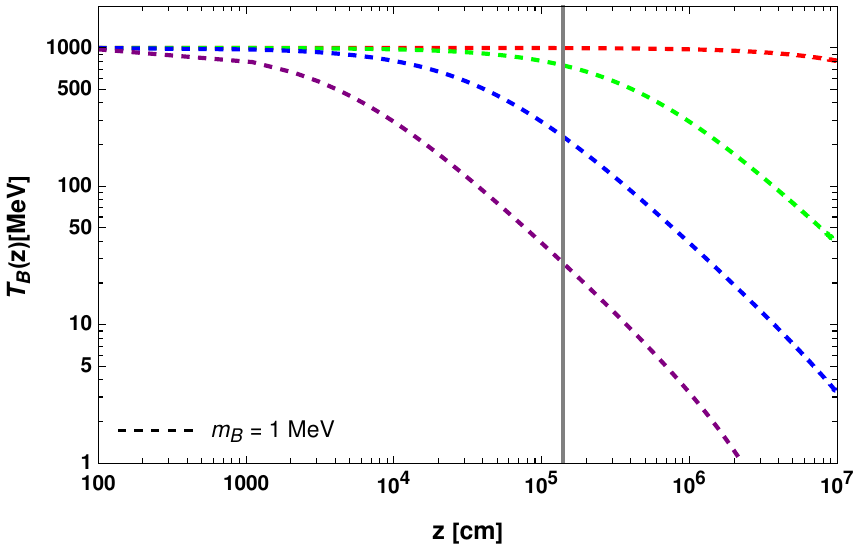}
\includegraphics[width=7cm,height=5cm]{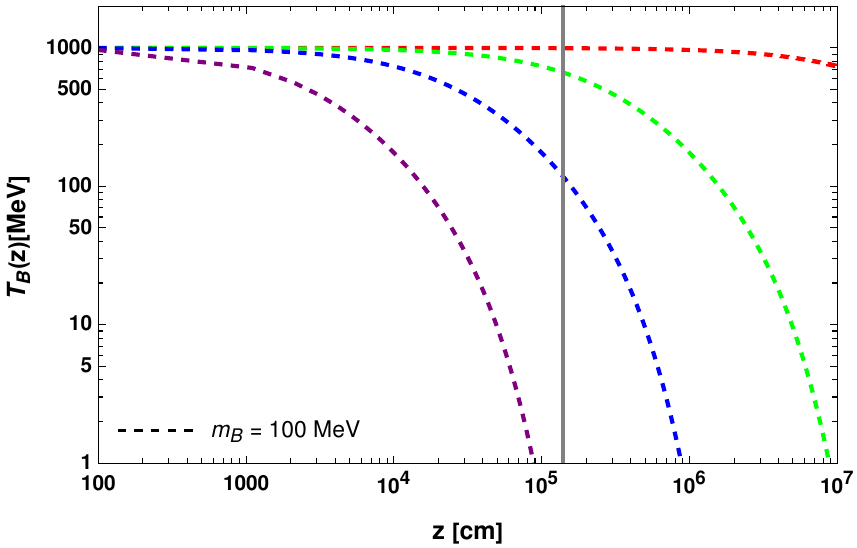}
\includegraphics[width=7cm,height=5cm]{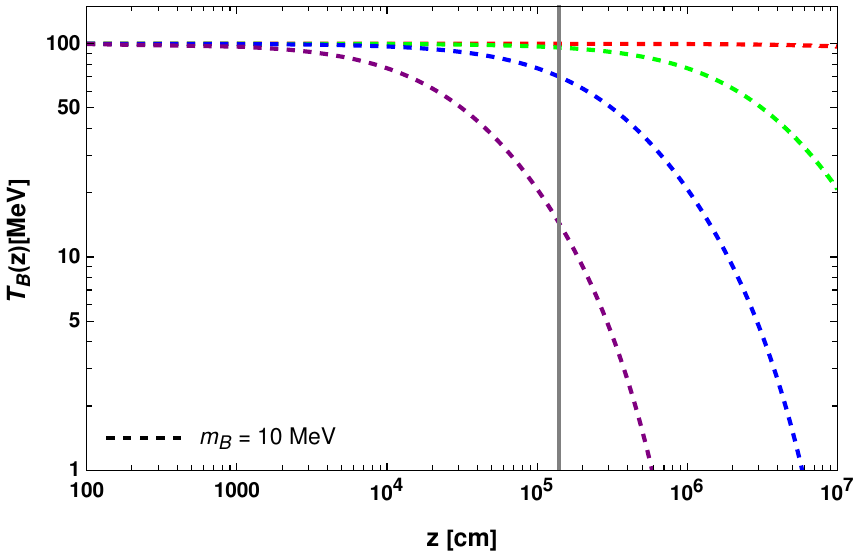}
\includegraphics[width=7cm,height=5cm]{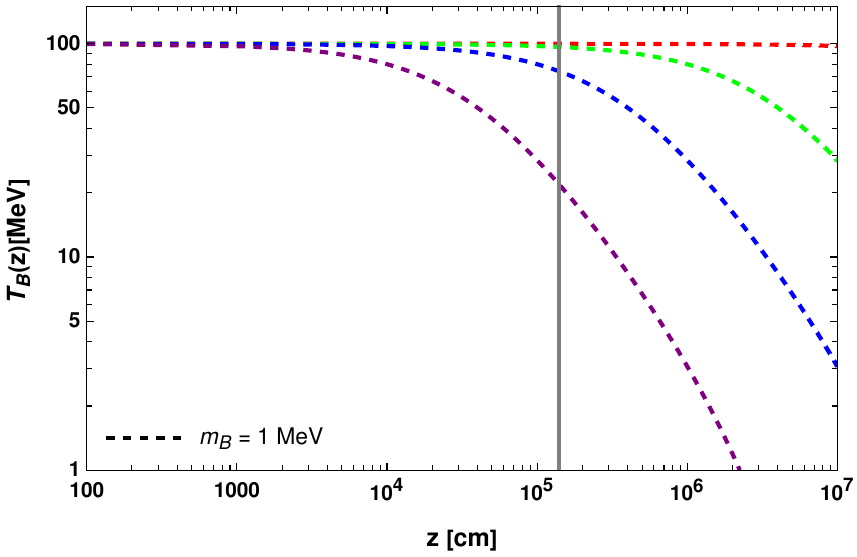}
\end{center}
\caption{Attenuation of kinetic energy of BDM as a function of distance ($z$) is shown when $F_N^2(q^2)= 1$ and for different DM-nucleon cross sections. $\sigma_{Bn} = 3\times10^{-32}$ cm$^2$ (red), $3\times10^{-30}$ cm$^2$ (green), $3\times10^{-29}$ cm$^2$ (blue), $3\times10^{-28}$ cm$^2$ (purple). The gray vertical line shows $z = 1.4$ km is the depth of XENON experiment.}
\label{Attn_plot2}
\end{figure}
In Fig.\ref{Attn_plot2} we show the attenuation effect for two different initial kinetic energies : 1 GeV and 100 MeV. If the mass of the DM is fixed, but the initial kinetic energies are different, then for larger kinetic energy the attenuation is more. For example, if the BDM has mass $\sim 100$ MeV and very large initial kinetic energy ($\sim 1$ GeV), then the final kinetic energy of the BDM reaching the underground detector will suffer from heavy attenuation if the scattering cross section is also large. However, for smaller scattering cross section the attenuation is comparatively less when other parameters are unchanged.
\subsection{Nuclear scattering: Form factor $\neq 1$}
\begin{figure}[htb!]
\begin{center}
\includegraphics[width=8cm,height=6cm]{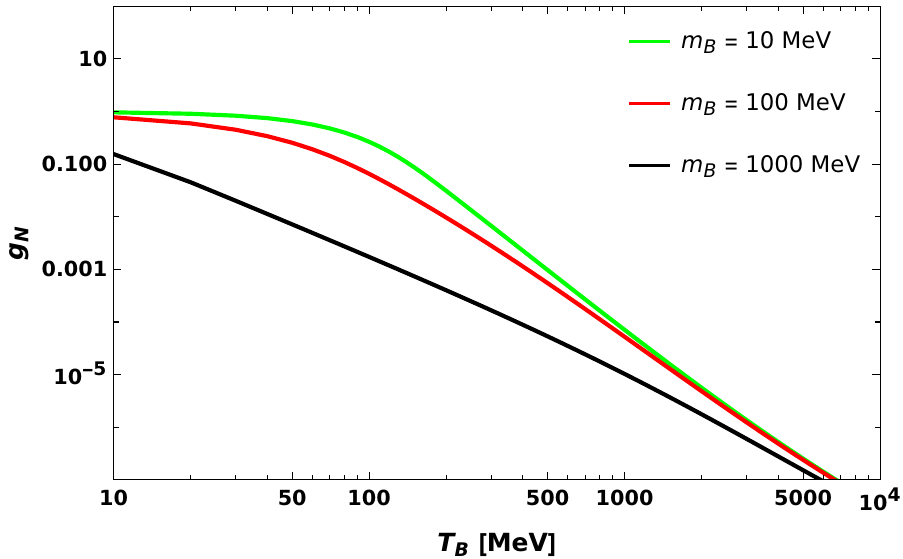}
\end{center}
\caption{Variation of $g_N$ with the kinetic energy of the boosted dark matter is shown for different boosted dark matter masses.}
\label{gN_plot}
\end{figure}

Boosted dark matter particles travel with high kinetic energy and hence the momentum transfer($q$) is large when the DM particles scatter with the nuclei. Hence, the effective DM-nucleus cross section depends on the form factor of the nuclei $F_N(q^2)$. In other words, the effect of the Helm form factor \cite{Helm:1956zz,Lewin:1995rx} becomes non-negligible for larger $q$ because $g_N (T_B)$ takes values less than $1$.
The Helm form factor can be approximated as a Gaussian form factor,
\begin{equation}
    {F_N(q^2)}\approx e^{-q^2/\Lambda_N^2} \quad (\text{for }q R_1
    < \zeta_1),
\end{equation}
where $\Lambda_N^{-2} \approx R_1^2/a^2+s^2/2$, $a\approx3.2$ and $\zeta_1$ the first zero of Bessel function ($\zeta_1$ = 4.449).  For oxygen nuclei $\Lambda_N\approx 0.207~\text{GeV}$.
The integration over the form factor in Eq. (\ref{gN}) results in, 
\begin{equation}
g_{N}(T_B)=\frac{\Lambda_N^2} {8 m_N^2 T_N^{max}(T_B)^2}\left(\Lambda_N^2-e^{-\frac{4 m_N T_N^{max}(T_B)}{\Lambda_N^2}} \left(\Lambda_N^2+4 m_N T_N^{max}(T_B)\right)\right)
\end{equation}
In Fig.\ref{gN_plot} we show the variation of $g_N$ with  kinetic energy ($T_B$) of the boosted dark matter particles.
When the mass of the boosted dark matter is 100 MeV or smaller, $g_N$ is almost constant ($\sim 1$) for kinetic energies in the range $T_B<100$ MeV. But when the kinetic energy is larger than 100 MeV, $g_N$ starts to decrease. This eventually increases the mean free path for the energy loss [see Eq. (\ref{1_by_ell})] and the kinetic energy does not suffer from large attenuation as compared to the case when $g_N=1$ ($F_N(q^2)\rightarrow 1$). Moreover, if the boosted dark matter is heavier than 100 MeV then $g_N$ is less than $1$ even at kinetic energies $\leq 10$ MeV.

\begin{figure}
\includegraphics[width=5.5cm,height=5cm]{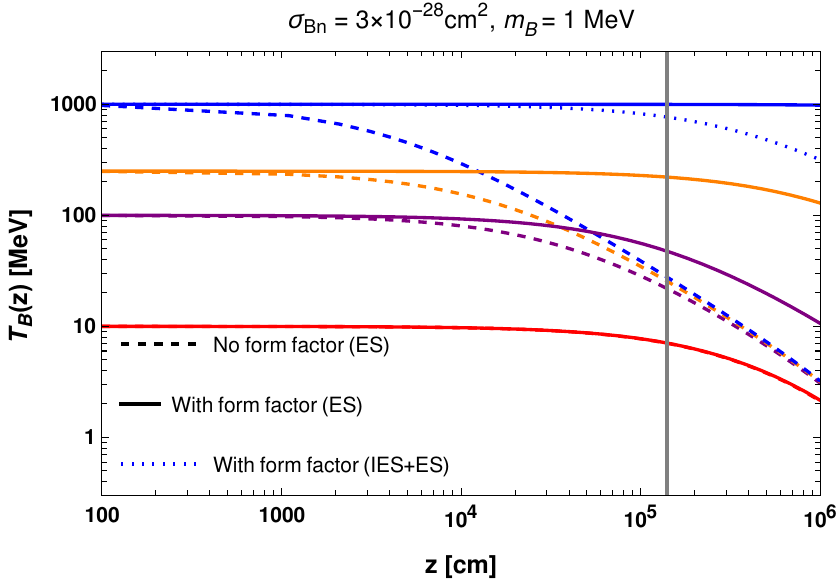}
\includegraphics[width=5.5cm,height=5cm]{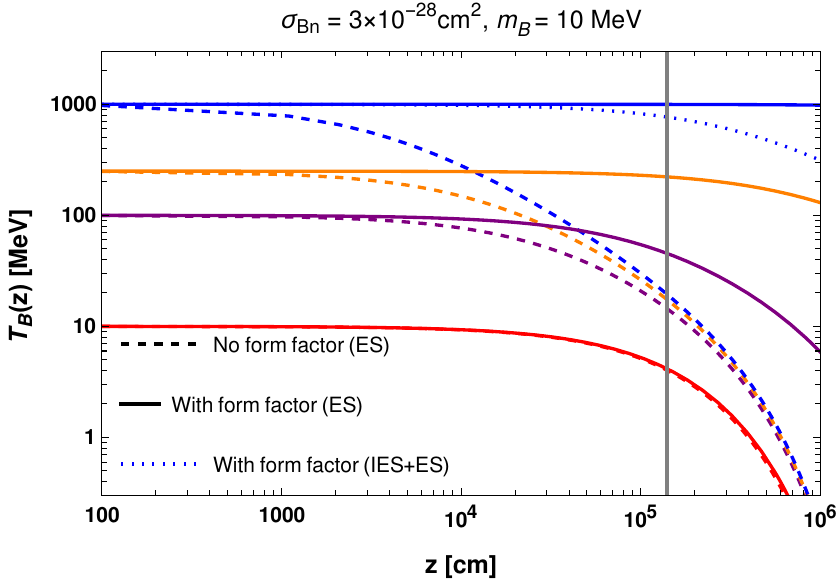}
\includegraphics[width=5.5cm,height=5cm]{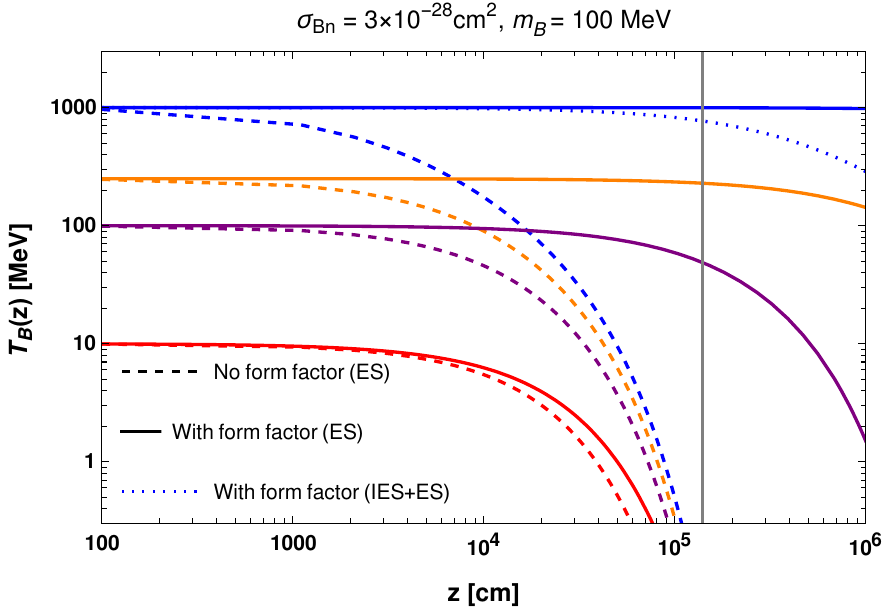}
\caption{Kinetic energy of the boosted dark matter as a function of $z$ is shown for different initial values of the kinetic energy. The boost factors ($m_A/m_B$) are $10-1000$ (left), $1-100$ (middle), $0.1-10$ (right). $\sigma_{Bn} = 3\times10^{-28}$ cm$^2$ and the gray vertical line shows $z = 1.4$ km  for the XENON experiment.}
\label{Full_nucl_attn}
\end{figure}

For $g_N<1$ the expression of kinetic energy as a function of $z$ [Eq. (\ref{eq:dTdz})] is not exactly solvable.
For kinetic energies more than 1 GeV, some approximate solutions exist (\cite{Xia:2021vbz}) but for kinetic energies less than 1 GeV, we solve Eq. (\ref{eq:dTdz}) numerically.
In Fig.\ref{Full_nucl_attn} we plot the numerical solution of the kinetic energy as a function of $z$ for different initial values. The mass of the boosted dark matter and DM-nucleon cross section are kept fixed. We also show the scenario with $F_N(q^2)=1$ by the dashed lines.
In the elastic scattering process involving BDM and the nucleus, the BDM particles interact with the nucleus as a unified entity when the initial kinetic energy is below 1 GeV. Conversely, if kinetic energy exceeds 1 GeV, inelastic scattering effects begin to play a more significant role. The maximum recoil energy deposited by a BDM with initial kinetic energy 1 GeV is around 700 MeV. At such high energy, due to large energy transfer, the BDM particle interacts directly with the constituent nucleons. Hence the inelastic scattering starts contributing at larger kinetic energies of the BDM. The dominating part of the inelastic scattering at this energy range is quasielastic scattering \cite{Alvey:2022pad}. For the attenuation scenarios with initial kinetic energy 1 GeV, we include the effect of inelastic scattering which is shown by the blue dotted line in Fig.\ref{Full_nucl_attn}.

If the kinetic energy of the BDM is 100 MeV or less, the numerical solution with form factor ($F_N^2(q^2)\neq 1$) almost agrees with the case when form factor is one ($F_N^2(q^2)=1$). 
When ($F_N^2(q^2)=1$) and the initial kinetic energy is between (100-1000 MeV), which corresponds to the boost factor $\sim 10^2$, the final kinetic energy suffers from significant attenuation.
But the effect of the form factor becomes more relevant at larger energies and with $F_N^2(q^2)\neq 1$, the kinetic energy of the BDM suffer from less attenuation (in some cases no attenuation at all), as shown by the solid lines in Fig.\ref{Full_nucl_attn}. For example, for 1 MeV BDM particle, if the initial kinetic energy is 100 MeV, the final kinetic energy at the XENON experiment (1.4 km underground) will be $\sim 50$ MeV. For experiments situated in more depth, $z> 1.4$ km, kinetic energy will be more attenuated. Increasing the cross section [even by $\mathcal{O}(1)$] results in significantly stronger attenuation. However, if only elastic scattering is considered, then the BDM with 1 GeV kinetic energy suffers almost no attenuation (solid blue line). But attenuation of the kinetic energy is observed if the effect of inelastic scattering is included (dotted blue lines). The attenuation in the kinetic energy is more as we increase the mass of the BDM to 10 MeV and 100 MeV but only for initial kinetic energy less than 100 MeV. We have also checked that decreasing the cross section shows very less to no significant effect when the initial kinetic energy is large ($\geq 100$ MeV).
The flux of the BDM in terms of the attenuated kinetic energy is expressed as, 
\begin{align}
   \frac{d \phi}{d T_B}\bigg\rvert_{z}=\frac{4m_B^2e^{z/\ell}}{\left(2m_B + T_B - T_B e^{z/\ell}\right)^2}\frac{d \phi}{d T_B}\bigg\rvert_{z=0}.
\end{align}
\begin{figure}
\includegraphics[width=5.7cm,height=5cm]{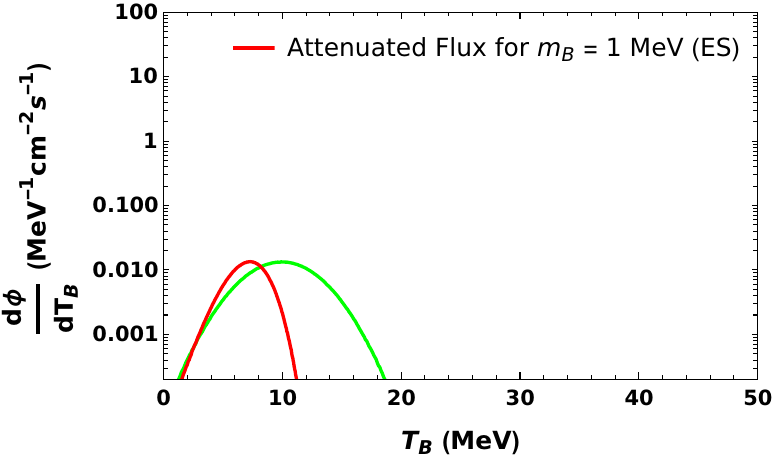}
\includegraphics[width=5.7cm,height=5cm]{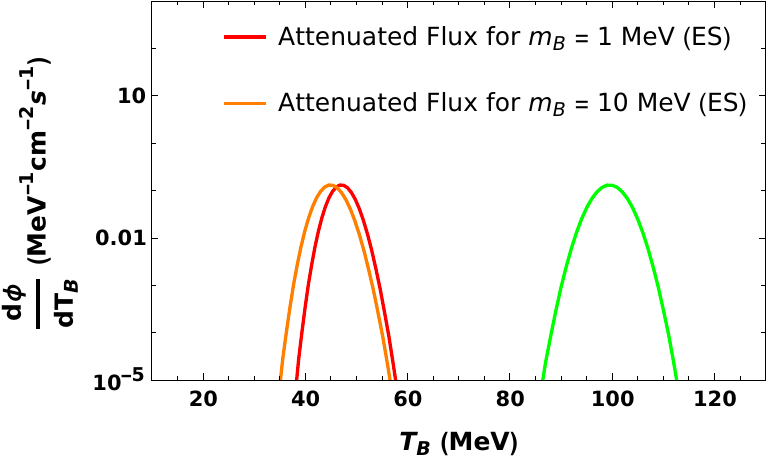}
\includegraphics[width=5.7cm,height=5cm]{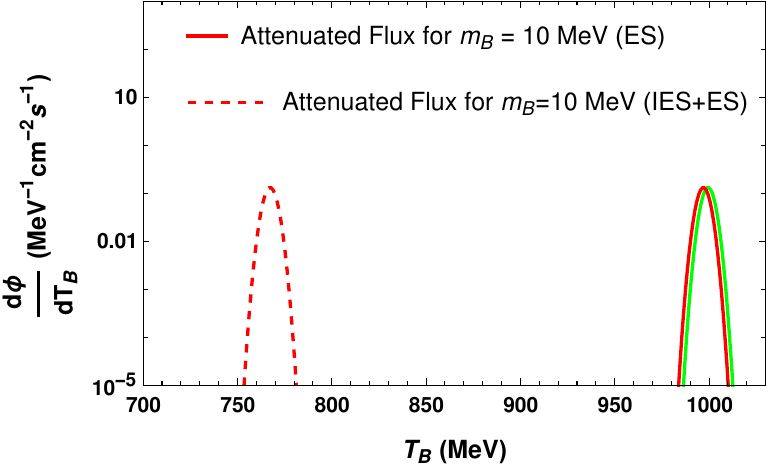}
\caption{Initial flux (green) and attenuated flux (red and orange) at $z=1.4$ km {\it with form factor} ($F_N^2(q^2)\neq 1$) for different $T_B=m_A$ and $m_B$. 
$\sigma_{Bn}$ is fixed at $3\times10^{-28}$ cm$^2$ and width of the BDM flux is $\sigma=3$.The boost is $=m_A/m_B$, and the peak value of $T_B$ is at $m_A$. For
$T_B\sim m_A=1000$ MeV, the effect of inelastic scattering is also considered.}
\label{Attnd_nucl_flux}
\end{figure}
The boosted dark matter flux reaching the XENON experiment is calculated using the form factors as discussed in the previous section.
In Fig.\ref{Attnd_nucl_flux} we show the boosted dark matter flux reaching the Earth with peaks at $T_B(=m_A)=10, 100$ and $1000$ MeV (in green color) 
as function of initial kinetic energy and the flux reaching the XENON experiment (in Red color) as a function of the attenuated kinetic energy. 
For the kinetic energy in the bottom axis we choose a common variable ($T_B$) to plot the original and attenuated flux and all distributions are normalized to 1. 
We observe a significant shift in the peak position of the distribution due to the attenuation of the kinetic energy. 
If the boosted dark matter mass is around $1$ MeV or less and the kinetic energy is $10$ MeV, the peak of the distribution does 
not shift much [Fig.\ref{Attnd_nucl_flux}(left)].
The shift is more for initial kinetic energy, $T_B\sim m_A= 100$ MeV [Fig.\ref{Attnd_nucl_flux}(middle)]. However at large kinetic energy, $T_B\sim m_A=1000$ MeV [Fig.\ref{Attnd_nucl_flux}(right)], the peak position does not shift at all 
if only elastic scattering is considered. In reality, at large kinetic energy the inelastic scattering also contributes and the position 
of the peak shifts even more (red dashed line) compared to the cases when $T_B\sim m_A=10,100$ MeV.  
\subsection{Scattering of BDM with electrons}
\begin{figure}[htb!]
\begin{center}
\includegraphics[width=7cm,height=5.5cm]{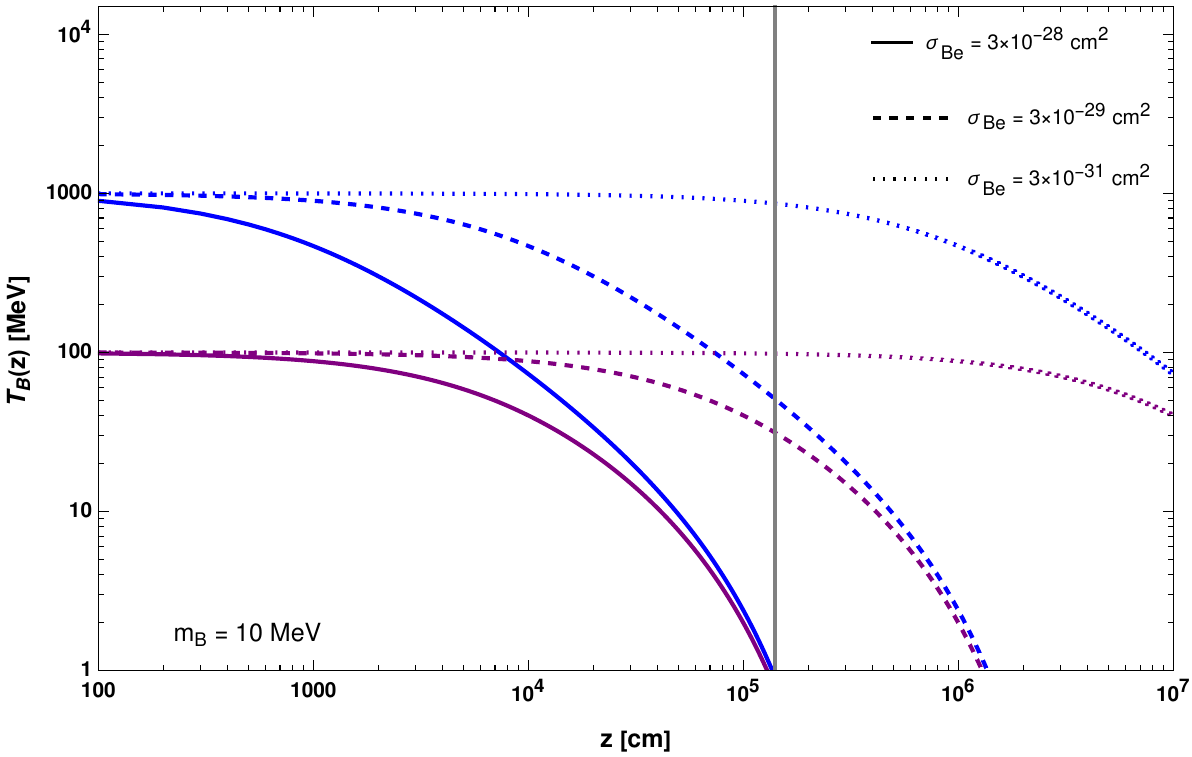}
\includegraphics[width=7cm,height=5.5cm]{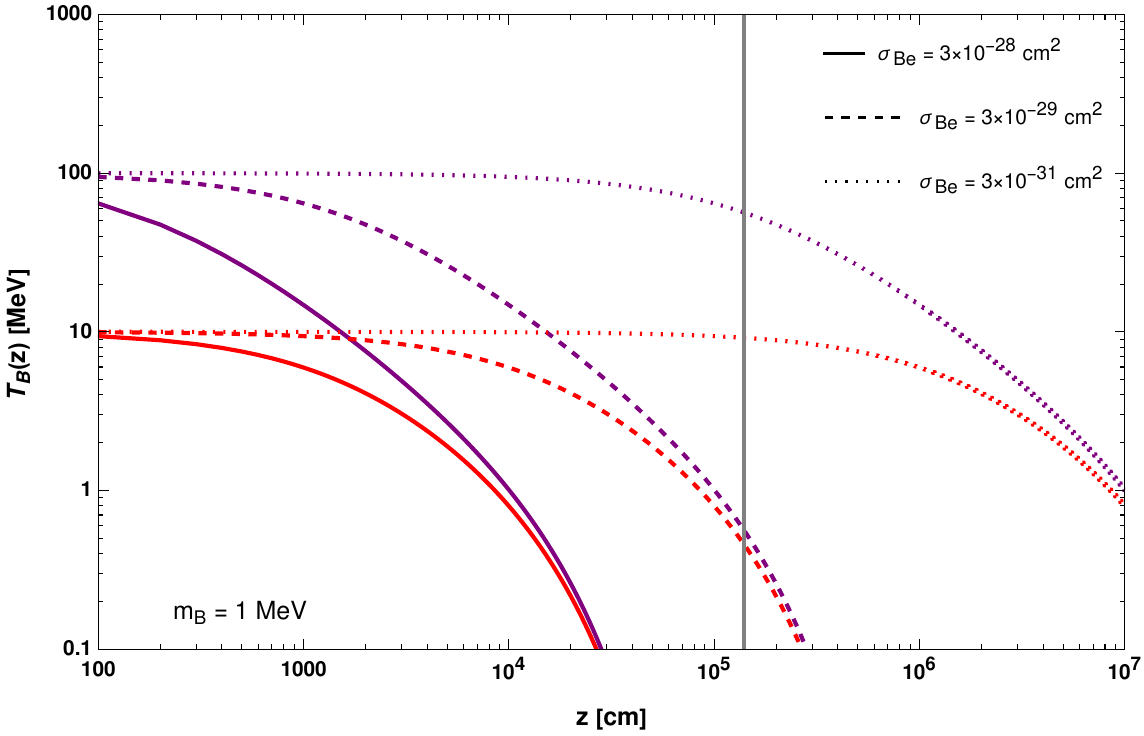}
\end{center}
\caption{Attenuation of kinetic energy for BDM-electron scattering for $m_B=10$ MeV(left) and $m_B=1$ MeV(right). The gray vertical line shows $z = 1.4$ km for the XENON experiment.}
\label{Full_elec_attn}
\end{figure}
Boosted dark matter traveling through the Earth's crust also interact with the electrons. Hence the kinetic energy suffers from attenuation due to the collisions with electrons. The analytical solution for attenuated kinetic energy is,
\begin{equation}\label{eq:Tz_noformfac}
	T_B(z)=\frac{T_B(0)
	e^{-z/\ell}}{1+\frac{T_B(0)}{2
	m_B}\left(1-e^{-z/\ell}\right)} ,
\end{equation}
same as in the case for DM-nucleus scattering. However, the mean free path for energy loss will be 
\begin{align}
\frac{1}{\ell_{E}}= n_e \sigma_{B e}\frac{2m_{e} m_{B}}{(m_{e}+m_{B})^2}, 
\end{align}
where $\sigma_{Be}$ is the DM-electron scattering cross section.
We explore the attenuation of the kinetic energy for different $\sigma_{Be}$ as shown in Fig.\ref{Full_elec_attn} for two different masses of the BDM, 10 MeV (left) and 1 MeV (right).
The scattering with the electrons (Fig.\ref{Full_elec_attn}) are found to be much stronger compared to the attenuation due to the DM-nucleus scattering (Fig.\ref{Full_nucl_attn}) in some scenarios. For example, in Fig.\ref{Full_elec_attn} (Left), if the BDM mass is 10 MeV and its kinetic energy is 100 MeV (boost$=10$), then for $\sigma_{Be}=3 \times 10^{-28}$cm$^2$, the final values of 
the kinetic energies are  $< 1 $ MeV.
Hence the attenuation due to electron scattering is very high for larger cross section, $\sigma_{Be} \geq 3 \times 10^{-28}$ cm$^2$.
Significant attenuation of the kinetic energy is seen when $\sigma_{Be} < 3 \times 10^{-28}$ cm$^2$. If $\sigma_{Be}\leq 3 \times 10^{-31}$ cm$^2$, 
the attenuation of the kinetic energy due to the scattering with the electrons is negligible at $z=1.4$ km.
\begin{figure}
\includegraphics[width=5.7cm,height=5cm]{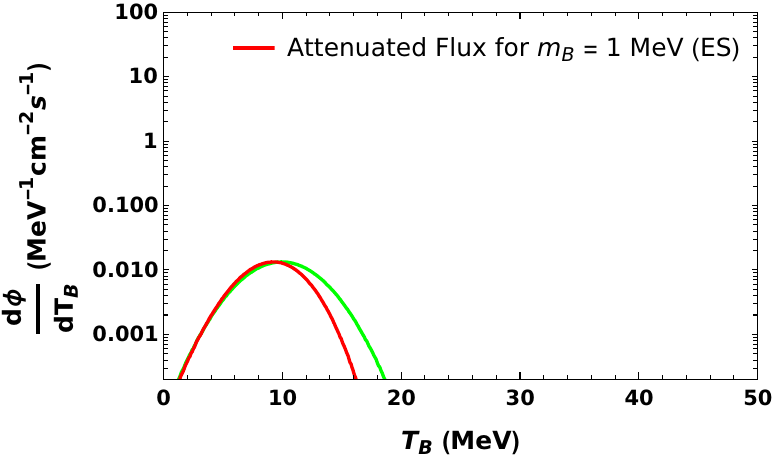}
\includegraphics[width=5.7cm,height=5cm]{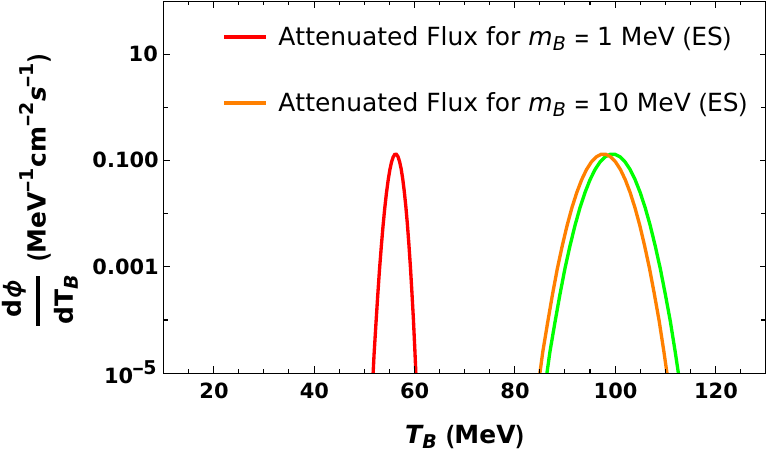}
\includegraphics[width=5.7cm,height=5cm]{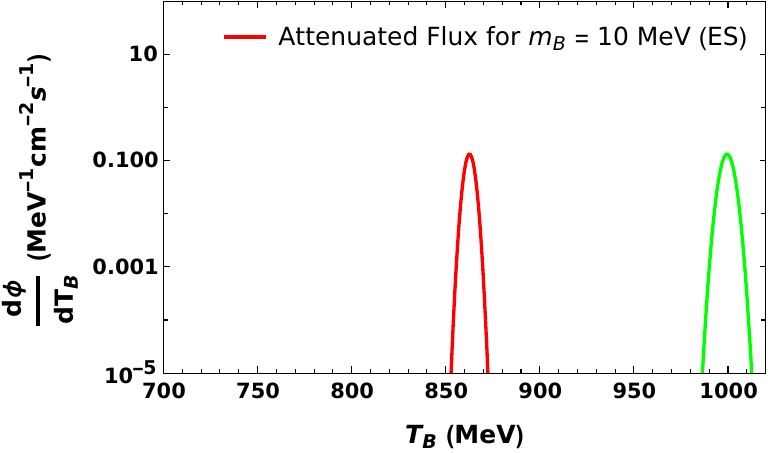}
\caption{Initial flux (green) and attenuated flux (red and orange) at $z=1.4$ km in electron scattering case for different $m_A$ and $m_B$. 
$\sigma_{Be}$ is fixed at $3\times10^{-31}$ cm$^2$ and width of the BDM flux is $\sigma=3$. The boost is $=m_A/m_B$, and the peak value of $T_B$ is at $m_A$.}
\label{Attnd_elec_flux}
\end{figure}
We also calculate the BDM flux reaching the XENON experiment
in Fig.\ref{Attnd_elec_flux} when the attenuation in the kinetic energy is not high, that is when $\sigma_{Be}=3 \times 10^{-31}$cm$^2$. Even with this small cross section, we observe a significant shift in the peak position of the BDM flux. It can be seen in Fig.\ref{Attnd_elec_flux} that when boost is $\mathcal{O}$(10$^2$)(right and middle), the peaks in initial and final flux are shifted further apart compared to the case when the boost is $\mathcal{O}(10)$ (left). For $\sigma_{Be}>3 \times 10^{-31}$cm$^2$ we observe that the attenuation of the kinetic energy is much stronger and the shift in the peak position is more. 
\subsection{Combined analysis}
\begin{figure}
\centering
\includegraphics[width=8cm,height=6cm]{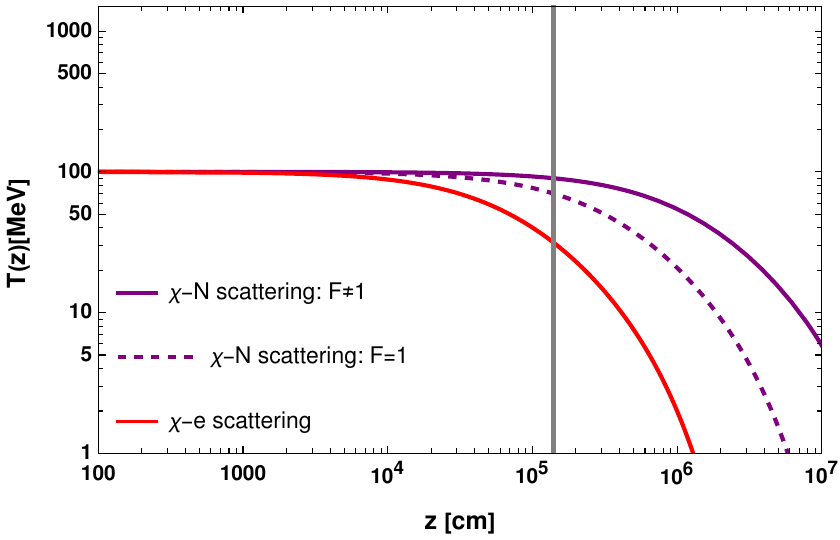}
\includegraphics[width=8cm,height=6cm]{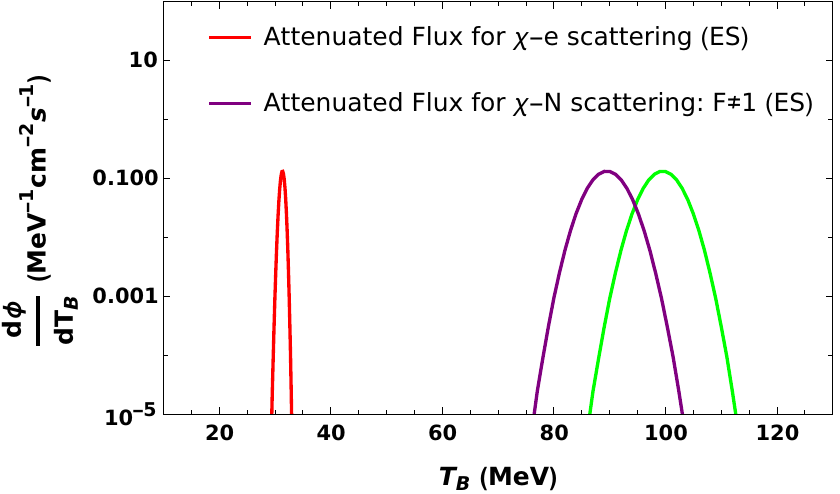}
\caption{Left: kinetic energy of the boosted dark matter as a function of $z$ is shown for DM-nucleus scattering (with and without form factor) and DM-electron scattering. $m_B$ = 10 MeV. The gray vertical line shows $z = 1.4$ km (XENON) and $\sigma_{Bn} = \sigma_{Be} = 3\times10^{-29}$ cm$^2$. Right: initial flux (green) and attenuated flux due to scattering with the nucleus (purple) and electron (red) scattering.}
\label{nucl_elec_att}
\end{figure}

To compare the effects of DM-nucleon and DM-electron scattering we show the attenuation in Fig.\ref{nucl_elec_att} assuming $\sigma_{Bn} = \sigma_{Be} = 3\times10^{-29}$ cm$^2$ and mass of the boosted dark matter at 10 MeV. The boosted dark matter has initial kinetic energy of 100 MeV, hence the boost factor is 10. We observe larger attenuation due to the scattering with the electrons than with the nucleons, resulting in a larger shift in the peak position in the former. Note that, only elastic scattering processes are contributing as kinetic energy is 100 MeV. For small scattering cross section, around $3\times10^{-31}$cm$^2$, attenuation due to both are found to be almost negligible. Hence we find a very interesting interplay between both scattering processes while observing the attenuation of the kinetic energy. 

One of the novelty of the boosted dark matter in the two component model is that the direct detection of the boosted dark matter ($B$)
gives insight about the indirect detection of the dominant dark matter candidate ($A$)\cite{Agashe:2014yua}, because the kinetic energy 
of the BDM is almost equal to the mass of the dominant DM component ($A$). But we observe that the position of the peak in the BDM flux shifts due to the attenuation of kinetic energy and it is not straight forward to associate the position of the peak to the mass of dark matter candidate $A$. The mass of the dominant dark matter $m_A$ is related to $T_B (Peak) - \delta T_B$, where $\delta T_B$ is the shift of the peak position in the BDM flux reaching the underground experiment. This shift is function of the masses of the DM, DM-nucleon scattering cross section and the width of the energy distribution.
 
\begin{figure}[htb!]
\centering
\includegraphics[width=8cm,height=6cm]{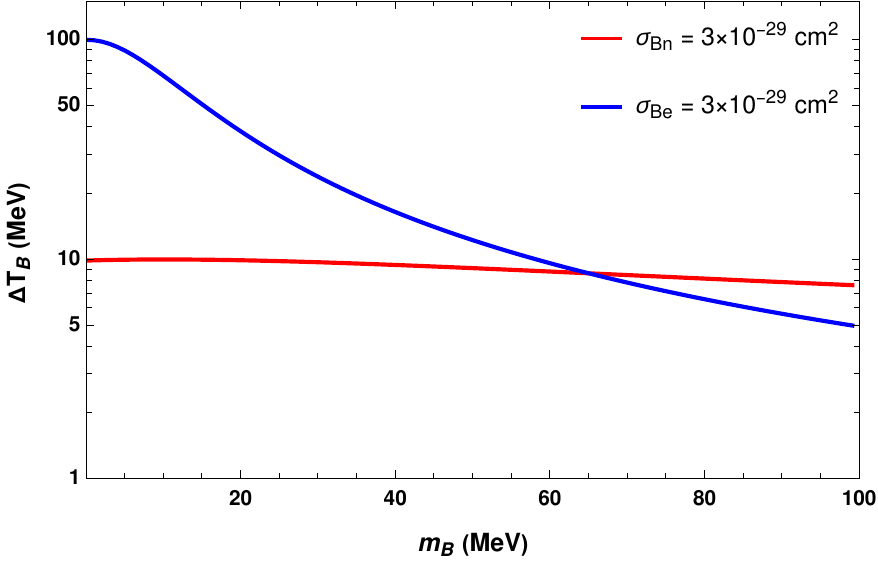}
\caption{Shift in the peak position after the attenuation in the kinetic energy at $z = 1.4$ km as a function of $m_B$. We assume the initial kinetic energy at $m_A=100$ MeV and  $\sigma_{Bn} = \sigma_{Be} = 3\times10^{-29}$ cm$^2$.}
\label{DeltaT}
\end{figure}

We plot the shift in the peak position as a function of the BDM mass, evaluated at $z = 1.4$ km in Fig.\ref{DeltaT}. We choose $\sigma_{Bn} = \sigma_{Be} = 3\times10^{-29}$ cm$^2$ and 
$m_A=100$ MeV. Hence the initial kinetic energy is 100 MeV as well. If we observe from left to right of Fig.\ref{DeltaT}, as the mass of the BDM increases, the boost decreases. Interestingly, in  the DM-nucleon scattering, the shift 
in the peak position remain almost constant. The shift is comparatively less at larger masses of the BDM or low boost ($\sim 1$). However, for the scattering with the 
electrons, the variation in $\delta T_B$ is more. Toward the left, if the mass of the BDM is small, and boost is large, a very large shift in the peak occurs. 
However, at large BDM masses or in other words, low boost region, the shift is very small. A small shift for the heavier BDM occurs because heavier dark matter particles get less attenuated than the lighter ones. Interestingly, if the BDM mass is more than 60 GeV, the effect of DM-nucleon scattering is more than the electron scattering. This is a very crucial observation because from Fig.\ref{Full_nucl_attn} and Fig.\ref{Attnd_elec_flux} we observe that the attenuation increases with the mass of BDM for the DM-nucleon scattering and decreases for the electron scattering. 
We have also checked that if $m_A=1000$ MeV, that is the kinetic energy is large, the conclusion remains almost the same.
\section{CONCLUSION}
Boosted dark matter, which is the subdominant component of dark matter may originate from the annihilation of the dominant dark matter 
species and the annihilation can be modeled in a two component dark matter scenario.  
Before reaching the underground direct detection experiments, boosted dark matter particles suffer from collision with the electron 
and the nucleus of the atoms present in Earth's atmosphere and its crust.
As kinetic energy experiences significant attenuation when passing through the Earth's crust, understanding this effect is crucial for establishing constraints on dark matter models based on findings from XENON, LZ, and other direct detection experiments.
We have examined the reduction in the kinetic energy of BDM particles within a general framework derived from two-component dark matter models. Our analysis demonstrates how this attenuation is influenced by various model specifics, including: (1) The masses of the dark matter components, which in turn affect the boost factor of the dark matter, (2) The annihilation cross section of the primary dark matter species present in the galactic halo, and (3) The scattering cross sections for interactions between dark matter and both electrons and nucleons.

To our knowledge this is the first attempt to address the attenuation of kinetic energy of the boosted dark matter in a two component scenario except \cite{Su:2023zgr}. 
The production of BDM in the two component model differs from the other sources of BDM in two aspects. First, the boost depends on the mass of the DM species, and second, 
the flux of the BDM is almost monoenergetic. Thus the shape of the BDM flux differs from the other scenarios such as DSNB boosted DM, CR boosted dark matter, 
BDM from PBH etc which has been studied vastly.  
We have performed a detailed analysis of the attenuation of the kinetic energy of BDM due to its scattering with the nucleus of the atoms present in the elements of Earth's crust. We have shown the effect considering the size of the nucleus by including the Helm form factor in our calculation. 
Moreover, we cross checked our analysis with ref:\cite{DeRomeri:2023ytt} and the results are in agreement. 
We find that for BDM masses in the range $1 - 100$ MeV,
if the initial kinetic energy is between $1-1000$ MeV, the attenuation in the kinetic energy is not negligible due to the DM-nucleon scattering.
For more accuracy, we include the effect of inelastic scattering when the kinetic energy is large.
We have studied the effect of DM-electron scattering as well and found that the attenuation is stronger for DM-electron
scattering compared to the DM-nucleon scattering, except when the boosted dark matter mass is large and the boost is small.
We also found that the attenuation of the kinetic energy is highly dependent on the DM-electron and DM-nucleon scattering cross sections.

We have analyzed the BDM with medium and high boost and we find that due to the attenuation of the kinetic energy of the BDM, the peak of the BDM flux shifts to a lower value. In the  
large boost region the shift is more. Hence, the peak value of the attenuated flux is not exactly at $m_A$. Thus, indirect detection of $m_A$ from the direct detection of $m_B$ is possible but 
one has to relate $m_A$ with the shift in the position of the peak and the change in the peak position depends on the model parameters.
We would like to mention that Monte Carlo (MC) simulations can yield a more precise estimate of the energy loss, specially when the boosted dark matter 
has very high kinetic energy and the scattering cross section is larger than $10^{-29}$cm$^{2}$. If numerical simulation is considered, then the attenuation 
is expected to be much stronger in some cases. This would result in larger shift in the peak position of the boosted dark matter flux. 
We plan to address this in a future work where we will place limits on the two component dark matter model from the possible detection 
of the boosted dark matter at future direct detection experiments.
\section*{ACKNOWLEDGMENTS}
The authors would like to thank M.R.Gangopadhyay for useful discussions. This work is supported fully by the Department of Science and Technology, Government of India under the SRG grant, Grant Agreement No. SRG/2022/000363 and CRG grant with Grant Agreement No. CRG/2022/004120.
\bibliographystyle{unsrt}
\bibliography{reference}

\begin{thebibliography}{10}

\bibitem{Bertone:2004pz}
Gianfranco Bertone, Dan Hooper, and Joseph Silk.
\newblock {Particle dark matter: Evidence, candidates and constraints}.
\newblock {\em Phys. Rept.}, 405:279--390, 2005.

\bibitem{Planck:2018vyg}
N.~Aghanim et~al.
\newblock {Planck 2018 results. VI. Cosmological parameters}.
\newblock {\em Astron. Astrophys.}, 641:A6, 2020.
\newblock [Erratum: Astron.Astrophys. 652, C4 (2021)].

\bibitem{XENON:2023cxc}
E.~Aprile et~al.
\newblock {First Dark Matter Search with Nuclear Recoils from the XENONnT
  Experiment}.
\newblock {\em Phys. Rev. Lett.}, 131(4):041003, 2023.

\bibitem{LZ:2022lsv}
J.~Aalbers et~al.
\newblock {First Dark Matter Search Results from the LUX-ZEPLIN (LZ)
  Experiment}.
\newblock {\em Phys. Rev. Lett.}, 131(4):041002, 2023.

\bibitem{Lai:2023qub}
Michela Lai.
\newblock {Recent results from DEAP-3600}.
\newblock {\em JINST}, 18(02):C02046, 2023.

\bibitem{CRESST:2019jnq}
A.~H. Abdelhameed et~al.
\newblock {First results from the CRESST-III low-mass dark matter program}.
\newblock {\em Phys. Rev. D}, 100(10):102002, 2019.

\bibitem{SuperCDMS:2020ymb}
D.~W. Amaral et~al.
\newblock {Constraints on low-mass, relic dark matter candidates from a
  surface-operated SuperCDMS single-charge sensitive detector}.
\newblock {\em Phys. Rev. D}, 102(9):091101, 2020.

\bibitem{EDELWEISS:2020fxc}
Q.~Arnaud et~al.
\newblock {First germanium-based constraints on sub-MeV Dark Matter with the
  EDELWEISS experiment}.
\newblock {\em Phys. Rev. Lett.}, 125(14):141301, 2020.

\bibitem{SENSEI:2020dpa}
Liron Barak et~al.
\newblock {SENSEI: Direct-Detection Results on sub-GeV Dark Matter from a New
  Skipper-CCD}.
\newblock {\em Phys. Rev. Lett.}, 125(17):171802, 2020.

\bibitem{Fayet:2004bw}
Pierre Fayet.
\newblock {Light spin 1/2 or spin 0 dark matter particles}.
\newblock {\em Phys. Rev. D}, 70:023514, 2004.

\bibitem{Boehm:2003hm}
C.~Boehm and Pierre Fayet.
\newblock {Scalar dark matter candidates}.
\newblock {\em Nucl. Phys. B}, 683:219--263, 2004.

\bibitem{Boehm:2002yz}
C.~Boehm, T.~A. Ensslin, and J.~Silk.
\newblock {Can Annihilating dark matter be lighter than a few GeVs?}
\newblock {\em J. Phys. G}, 30:279--286, 2004.

\bibitem{Xia:2021vbz}
Chen Xia, Yan-Hao Xu, and Yu-Feng Zhou.
\newblock {Production and attenuation of cosmic-ray boosted dark matter}.
\newblock {\em JCAP}, 02(02):028, 2022.

\bibitem{Jho:2021rmn}
Yongsoo Jho, Jong-Chul Park, Seong~Chan Park, and Po-Yan Tseng.
\newblock {Cosmic-Neutrino-Boosted Dark Matter ($\nu$BDM)}.
\newblock 1 2021.

\bibitem{Yin:2018yjn}
Wen Yin.
\newblock {Highly-boosted dark matter and cutoff for cosmic-ray neutrinos
  through neutrino portal}.
\newblock {\em EPJ Web Conf.}, 208:04003, 2019.

\bibitem{Chao:2021orr}
Wei Chao, Tong Li, and Jiajun Liao.
\newblock {Connecting Primordial Black Hole to boosted sub-GeV Dark Matter
  through neutrino}.
\newblock 8 2021.

\bibitem{DeRomeri:2023ytt}
Valentina De~Romeri, Anirban Majumdar, Dimitrios~K. Papoulias, and Rahul
  Srivastava.
\newblock {XENONnT and LUX-ZEPLIN constraints on DSNB-boosted dark matter}.
\newblock {\em JCAP}, 03:028, 2024.

\bibitem{Das:2024ghw}
Anirban Das, Tim Herbermann, Manibrata Sen, and Volodymyr Takhistov.
\newblock {Energy-dependent boosted dark matter from diffuse supernova neutrino
  background}.
\newblock {\em JCAP}, 07:045, 2024.

\bibitem{Granelli:2022ysi}
Alessandro Granelli, Piero Ullio, and Jin-Wei Wang.
\newblock {Blazar-boosted dark matter at Super-Kamiokande}.
\newblock {\em JCAP}, 07(07):013, 2022.

\bibitem{Basu:2023wgo}
Arindam Basu, Amit Chakraborty, Nilanjana Kumar, and Soumya Sadhukhan.
\newblock {Viability of Boosted Light Dark Matter in a Two-Component Scenario}.
\newblock 10 2023.

\bibitem{Li:2023fzv}
Jinmian Li, Takaaki Nomura, Junle Pei, Xiangwei Yin, and Cong Zhang.
\newblock {Boosting indirect detection of a secluded dark matter sector}.
\newblock {\em Phys. Rev. D}, 108(3):035021, 2023.

\bibitem{Herrera:2021puj}
Gonzalo Herrera and Alejandro Ibarra.
\newblock {Direct detection of non-galactic light dark matter}.
\newblock {\em Phys. Lett. B}, 820:136551, 2021.

\bibitem{Herrera:2023fpq}
Gonzalo Herrera, Alejandro Ibarra, and Satoshi Shirai.
\newblock {Enhanced prospects for direct detection of inelastic dark matter
  from a non-galactic diffuse component}.
\newblock {\em JCAP}, 04:026, 2023.

\bibitem{DarkSide-20k:2017zyg}
C.~E. Aalseth et~al.
\newblock {DarkSide-20k: A 20 tonne two-phase LAr TPC for direct dark matter
  detection at LNGS}.
\newblock {\em Eur. Phys. J. Plus}, 133:131, 2018.

\bibitem{PandaX:2014mem}
XiGuang Cao et~al.
\newblock {PandaX: A Liquid Xenon Dark Matter Experiment at CJPL}.
\newblock {\em Sci. China Phys. Mech. Astron.}, 57:1476--1494, 2014.

\bibitem{Chen:2021ifo}
Yifan Chen, Bartosz Fornal, Pearl Sandick, Jing Shu, Xiao Xue, Yue Zhao, and
  Junchao Zong.
\newblock {Earth shielding and daily modulation from electrophilic boosted dark
  particles}.
\newblock {\em Phys. Rev. D}, 107(3):033006, 2023.

\bibitem{Herbermann:2024kcy}
Tim Herbermann, Manfred Lindner, and Manibrata Sen.
\newblock {Attenuation of Cosmic Ray Electron Boosted Dark Matter}.
\newblock 8 2024.

\bibitem{Agashe:2014yua}
Kaustubh Agashe, Yanou Cui, Lina Necib, and Jesse Thaler.
\newblock {(In)direct Detection of Boosted Dark Matter}.
\newblock {\em JCAP}, 10:062, 2014.

\bibitem{Borah:2021yek}
Debasish Borah, Manoranjan Dutta, Satyabrata Mahapatra, and Narendra Sahu.
\newblock {Boosted self-interacting dark matter and XENON1T excess}.
\newblock {\em Nucl. Phys. B}, 979:115787, 2022.

\bibitem{Borah:2021jzu}
Debasish Borah, Manoranjan Dutta, Satyabrata Mahapatra, and Narendra Sahu.
\newblock {Muon $(g-2)$ and XENON1T excess with boosted dark matter in
  $L_{\mu}-L_{\tau}$ model}.
\newblock {\em Phys. Lett. B}, 820:136577, 2021.

\bibitem{Starkman:1990nj}
Glenn~D. Starkman, Andrew Gould, Rahim Esmailzadeh, and Savas Dimopoulos.
\newblock {Opening the Window on Strongly Interacting Dark Matter}.
\newblock {\em Phys. Rev. D}, 41:3594, 1990.

\bibitem{Kouvaris:2014lpa}
Chris Kouvaris and Ian~M. Shoemaker.
\newblock {Daily modulation as a smoking gun of dark matter with significant
  stopping rate}.
\newblock {\em Phys. Rev. D}, 90:095011, 2014.

\bibitem{Helm:1956zz}
Richard~H. Helm.
\newblock {Inelastic and Elastic Scattering of 187-Mev Electrons from Selected
  Even-Even Nuclei}.
\newblock {\em Phys. Rev.}, 104:1466--1475, 1956.

\bibitem{Su:2022wpj}
Liangliang Su, Lei Wu, Ning Zhou, and Bin Zhu.
\newblock {Accelerated-light-dark-matter\textendash{}Earth inelastic scattering
  in direct detection}.
\newblock {\em Phys. Rev. D}, 108(3):035004, 2023.

\bibitem{Su:2023zgr}
Liangliang Su, Lei Wu, and Bin Zhu.
\newblock {An improved bound on accelerated light dark matter}.
\newblock {\em Sci. China Phys. Mech. Astron.}, 67(2):221012, 2024.

\bibitem{PandaX:2023tfq}
Xuyang Ning et~al.
\newblock {Search for Light Dark Matter from the Atmosphere in PandaX-4T}.
\newblock {\em Phys. Rev. Lett.}, 131(4):041001, 2023.

\bibitem{Alvey:2022pad}
James Alvey, Torsten Bringmann, and Helena Kolesova.
\newblock {No room to hide: implications of cosmic-ray upscattering for
  GeV-scale dark matter}.
\newblock {\em JHEP}, 01:123, 2023.

\bibitem{Bhattacharya:2013hva}
Subhaditya Bhattacharya, Aleksandra Drozd, Bohdan Grzadkowski, and Jose Wudka.
\newblock {Two-Component Dark Matter}.
\newblock {\em JHEP}, 10:158, 2013.

\bibitem{Guha:2024mjr}
Atanu Guha and Jong-Chul Park.
\newblock {Constraints on cosmic-ray boosted dark matter with realistic cross
  section}.
\newblock {\em JCAP}, 07:074, 2024.

\bibitem{Navarro:1995iw}
Julio~F. Navarro, Carlos~S. Frenk, and Simon D.~M. White.
\newblock {The Structure of cold dark matter halos}.
\newblock {\em Astrophys. J.}, 462:563--575, 1996.

\bibitem{1965TrAlm...5...87E}
J.~{Einasto}.
\newblock {On the Construction of a Composite Model for the Galaxy and on the
  Determination of the System of Galactic Parameters}.
\newblock {\em Trudy Astrofizicheskogo Instituta Alma-Ata}, 5:87--100, January
  1965.

\bibitem{Burkert:1995yz}
A.~Burkert.
\newblock {The Structure of dark matter halos in dwarf galaxies}.
\newblock {\em Astrophys. J. Lett.}, 447:L25, 1995.

\bibitem{Cirelli:2010xx}
Marco Cirelli, Gennaro Corcella, Andi Hektor, Gert Hutsi, Mario Kadastik, Paolo
  Panci, Martti Raidal, Filippo Sala, and Alessandro Strumia.
\newblock {PPPC 4 DM ID: A Poor Particle Physicist Cookbook for Dark Matter
  Indirect Detection}.
\newblock {\em JCAP}, 03:051, 2011.
\newblock [Erratum: JCAP 10, E01 (2012)].

\bibitem{Bhattacharjee:2012xm}
Pijushpani Bhattacharjee, Soumini Chaudhury, Susmita Kundu, and Subhabrata
  Majumdar.
\newblock {Sizing-up the WIMPs of Milky Way : Deriving the velocity
  distribution of Galactic Dark Matter particles from the rotation curve data}.
\newblock {\em Phys. Rev. D}, 87:083525, 2013.

\bibitem{Emken:2018run}
Timon Emken and Chris Kouvaris.
\newblock {How blind are underground and surface detectors to strongly
  interacting Dark Matter?}
\newblock {\em Phys. Rev. D}, 97(11):115047, 2018.

\bibitem{Ema:2018bih}
Yohei Ema, Filippo Sala, and Ryosuke Sato.
\newblock {Light Dark Matter at Neutrino Experiments}.
\newblock {\em Phys. Rev. Lett.}, 122(18):181802, 2019.

\bibitem{Wu:2022jln}
Ke-Yun Wu and Zhao-Hua Xiong.
\newblock {Spin-Dependent Scattering of Scalar and Vector Dark Matter on the
  Electron}.
\newblock {\em Symmetry}, 14(5):1061, 2022.

\bibitem{Lewin:1995rx}
J.~D. Lewin and P.~F. Smith.
\newblock {Review of mathematics, numerical factors, and corrections for dark
  matter experiments based on elastic nuclear recoil}.
\newblock {\em Astropart. Phys.}, 6:87--112, 1996.

\end{thebibliography}

\end{document}